\documentclass[fleqn,usenatbib]{mnras}

\usepackage{newtxtext,newtxmath}

\usepackage[T1]{fontenc}

\DeclareRobustCommand{\VAN}[3]{#2}
\let\VANthebibliography\thebibliography
\def\thebibliography{\DeclareRobustCommand{\VAN}[3]{##3}\VANthebibliography}

\usepackage{graphicx}	
\usepackage{tabularx}   
\usepackage{soul}       

\title[The cyclic behaviour in the N-S asymmetry]{The cyclic behaviour in the N-S asymmetry of sunspots and solar plages for the period 1910 to 1937 using data from Ebro catalogues} 

\author[V. de Paula et al.]{
V. de Paula,$^{1}$\thanks{E-mail: vdepaula@obsebre.es}
J.J. Curto,$^{1}$
and R. Oliver$^{2,3}$
\\
$^{1}$Observatori de l’Ebre (OE), CSIC-URL, Horta Alta 38, 43520 Roquetes, Spain.\\
$^{2}$Departament de Física, Universitat de les Illes Balears (UIB), 07122 Palma de Mallorca, Spain.\\
$^{3}$Institute of Applied Computing and Community Code (IAC3), Universitat de les Illes Balears (UIB), 07122 Palma de Mallorca, Spain.
}

\date{Accepted XXX. Received YYY; in original form ZZZ}

\pubyear{2021}

\begin{document}
\label{firstpage}
\pagerange{\pageref{firstpage}--\pageref{lastpage}}
\maketitle

\begin{abstract}
The heliophysics catalogues published by the Ebro Observatory during 1910--1937 have been converted into a digital format in order to provide the data for computational processing. This has allowed us to study in detail the North-South (N-S) asymmetry of solar activity in that period, focusing on two different structures located at two different layers of the solar atmosphere: sunspots (Photosphere) and solar plages (Chromosphere). The examination of the absolute and normalised N-S asymmetry indices in terms of their monthly sum of occurrences and areas has made possible to find out a cyclic behaviour in the solar activity, in which the preferred hemisphere changes systematically with a global period of $7.9 \pm 0.2~yr$. In order to verify and quantify accurately this periodicity and study its prevalence in time, we employed the RGO-USAF/NOAA sunspot data series during 1874--2016. Then, we examined each absolute asymmetry index time series through different techniques as the power spectrum analysis, the Complete Ensemble Empirical Mode Decomposition With Adaptive Noise algorithm, or the Morlet wavelet transform. The combined results reveal a cyclic behaviour at different time scales, consisting in two quite stable periodicities of $1.47 \pm 0.02~yr$ and $3.83 \pm 0.06~yr$, which coexist with another three discontinuous components with more marked time-varying periods with means of $5.4 \pm 0.2~yr$, $9.0 \pm 0.2~yr$, and $12.7 \pm 0.3~yr$. Moreover, during 1910--1937, only two dominant signals with averaged periods of $4.10 \pm 0.04~yr$ and $7.57 \pm 0.03~yr$ can be clearly observed. Finally, in both signals, periods are slightly longer for plages in comparison with sunspots.
\end{abstract}


\setcitestyle {aysep={,}}

\section{Introduction}

The search for periodicities in the solar activity has been the focus of many studies in the history of Solar Physics. The development of this issue led to the discovery of fundamental solar phenomena as the Spörer’s law \citep{Carrington1863} or the 11-yr solar cycle itself \citep{Schwabe1844}. 

There are strong evidences that suggest that the solar activity presents systematic variations with periods of the order of several solar cycles. Examples of these periodic variations, which seem to modulate the solar activity in the long-term and even could affect the Earth climate by producing unusual periods as the Maunder minimum \citep{KERN2012,RUZMAIKIN2015}, are the 87-yr Gleissberg cycle \citep{Gleissberg1939}, the 207-yr Suess-de Vries cycle \citep{suess_1980} or the 2400-yr Hallstatt cycle \citep{Vasiliev2002}. 

Periodic variations in the solar activity on time scales shorter than the 11-years Schwabe cycle have also been found. Two well-known examples of these short-term variations are the so-called Rieger periodicity of 154-158 days \citep{Rieger1984}, found especially in epochs of maximum activity \citep{Lean1990, Oliver1998, Krivova2002, Roy2020}; and the 1-4 yr quasi-biennial oscillations (QBOs) \citep{Benevolenskaya1995, Mursula2003, Bazilevskaya2014, Chowdhury2016, Chowdhury2019}.

The particular case of the North-South (N-S) asymmetry of solar activity has also been studied by looking for periodic behaviours. It has been found that the N-S asymmetry is modulated in the long-term, with an associated period of 8-12 solar cycles \citep{Waldmeier1957,Vizoso1990, Verma1993, Oliver1994, Pulkkinen1999, Li2002, Li2009, Knaack2004, Zhang2015}, as well as in a time scale of the order of the 11-years solar cycle \citep{Carbonell1993, Oliver1994,Oliver1996, Li2002, Knaack2004}, and even in shorter time scales, as 8.65 yr, 4.7 yr or 1.44 yr \citep{Ballester2005,Chowdhury2019, Ravindra2021}. 

The aim of this paper is, first, to prove that, at least during 1910—1937, there is a systematic change in the preferred hemisphere of solar activity, with a global period of approximately 7.5-8 yr. This is done by applying different techniques based on the power spectrum analysis, the Complete Ensemble Empirical Mode Decomposition With Adaptive Noise (CEEMDAN) method and the Morlet wavelet transform to the N-S asymmetry data derived from the sunspot data and Ca II-K solar plage series of records provided by the historical Ebro Observatory (EO) heliophysics catalogues in the period 1910—1937 and the sunspot data belonging to the Royal Greenwich Observatory (RGO) – United States Air Force (USAF) / National Oceanic and Atmospheric Administration (NOAA) in the period 1874—2016. The second aim is to characterise the evolution in time of this phenomenon and its coexistence with other periodicities at different timescales, not only in terms of the two features considered (monthly sum of occurrences and areas), but also in terms of the different layers of the solar atmosphere, by using sunspots (Photosphere) and solar plages (Chromosphere) as tracers. 

\section{Data}

The data used in this work are the monthly sum of occurrences and areas belonging to the EO series of sunspot and Ca II-K solar plage groups (hereafter solar plage groups) records in the period 1910--1937 (both inclusive), which have a temporal coverage of 81\%. This provided us with 672 rows of data entries belonging to 4212 sunspot groups and 5781 solar plage groups \citep {dePaula2020}. Both series of records can be found in the EO website: \url{http://www.obsebre.es/en/observatori-publications}. Moreover, in order to compare and test our data, and also expand our temporal window of study, we also employed the 1710 rows of entries corresponding to the monthly sunspot data recorded by the RGO from May of 1874 to the end of 1976 and by the USAF and the NOAA from 1977 to October of 2016, which are available in the website of the Solar Physics Group at NASA’s Marshall Space Flight Center: \url{https://solarscience.msfc.nasa.gov/greenwch.shtml}. 

All the sunspot and solar plage data have been analysed by focusing on the N-S asymmetry of the solar activity. To this end, the monthly sum of occurrences and areas of both structures have been calculated considering the two solar hemispheres separately.

\subsection{Ebro Observatory sunspot and solar plage records}
\label{sec:EOdata} 

\begin{figure*}
	\includegraphics[width=\textwidth]{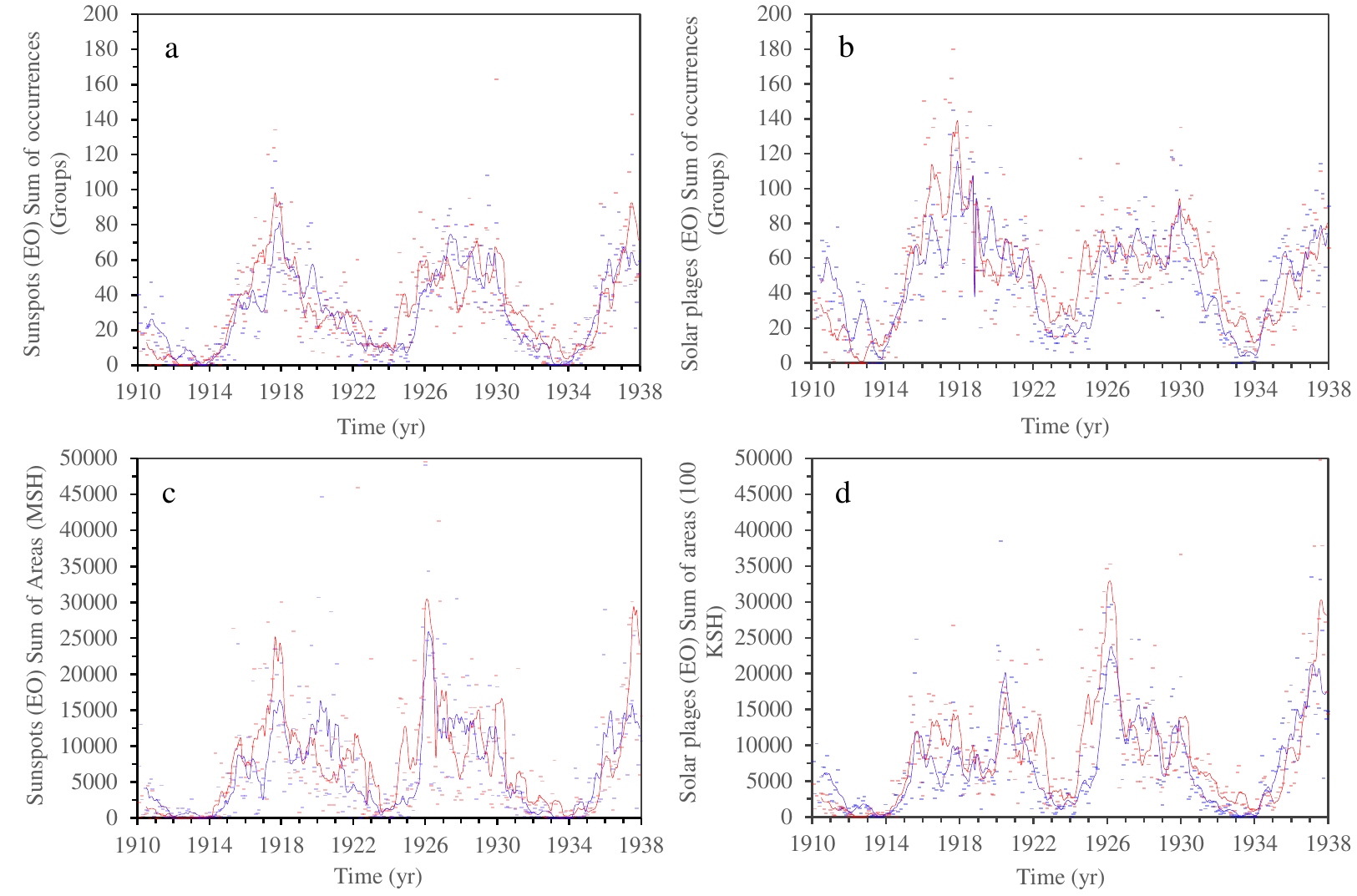}
    \caption{Evolution over time in the period 1910—1937 of the monthly sum of occurrences (upper panels \textit{a} and \textit{b}, respectively) and areas (lower panels \textit{c} and \textit{d}, respectively) of sunspot and solar plage groups. All the data correspond to the measurements performed at EO. Pale red and blue dots represent respectively, the monthly data corresponding to the northern and southern hemispheres. Red and blue solid lines are related to the respective 7-month running averages.}
    \label{fig:Figure1}
\end{figure*}

Figure \ref{fig:Figure1} shows a comparison between sunspots and solar plages groups in the total number of occurrences recorded in the EO at the northern and southern solar hemispheres during 1910--1937 (upper panels \textit{a} and \textit{b}), and their sum of areas (lower panels \textit{c} and \textit{d}), expressed in millionths of a solar hemisphere (MSH) in the case of sunspots, and in hundreds thousands of a solar hemisphere (100 KSH) in the case of solar plages.

The different integrals of each of the above-mentioned properties over the whole considered period reveal that in both structures, the northern hemisphere presents, on average, values about $5.7 \pm 0.6 \%$ higher than the southern. Moreover, on average the northern hemisphere also dominates the southern during the $55 \pm 1 \%$ of the studied period. Similar results were found in \citep{SEGUI2019, dePaula2020} examining the occurrence daily rate of the different morphologies of sunspots and solar plages groups. In addition, phase shifts between hemispheres can be noted in all panels every 3-5 yr. This asynchrony in the solar activity between the northern and southern hemispheres in terms of the total number of sunspot and solar plage groups occurrences and of their sum of areas implies that the solar activity starts before in one hemisphere than in the other \citep {Yi1992}, and could be related to the N-S asymmetry with a connection to the variation of the 11-years Schwabe cycle \citep{Temmer2006, Zolotova2006, McIntosh_2014, Deng2016, Schussler2018, Chowdhury2019, Ravindra2021}, or even to differences in the behaviour and field strength of the dynamo processes operating in both hemispheres \citep{Chowdhury2019, Ravindra2021}. 

\subsection{RGO-USAF/NOAA sunspot records}
\label{sec:RGOdata} 

\begin{figure*}
	\includegraphics[width=\textwidth]{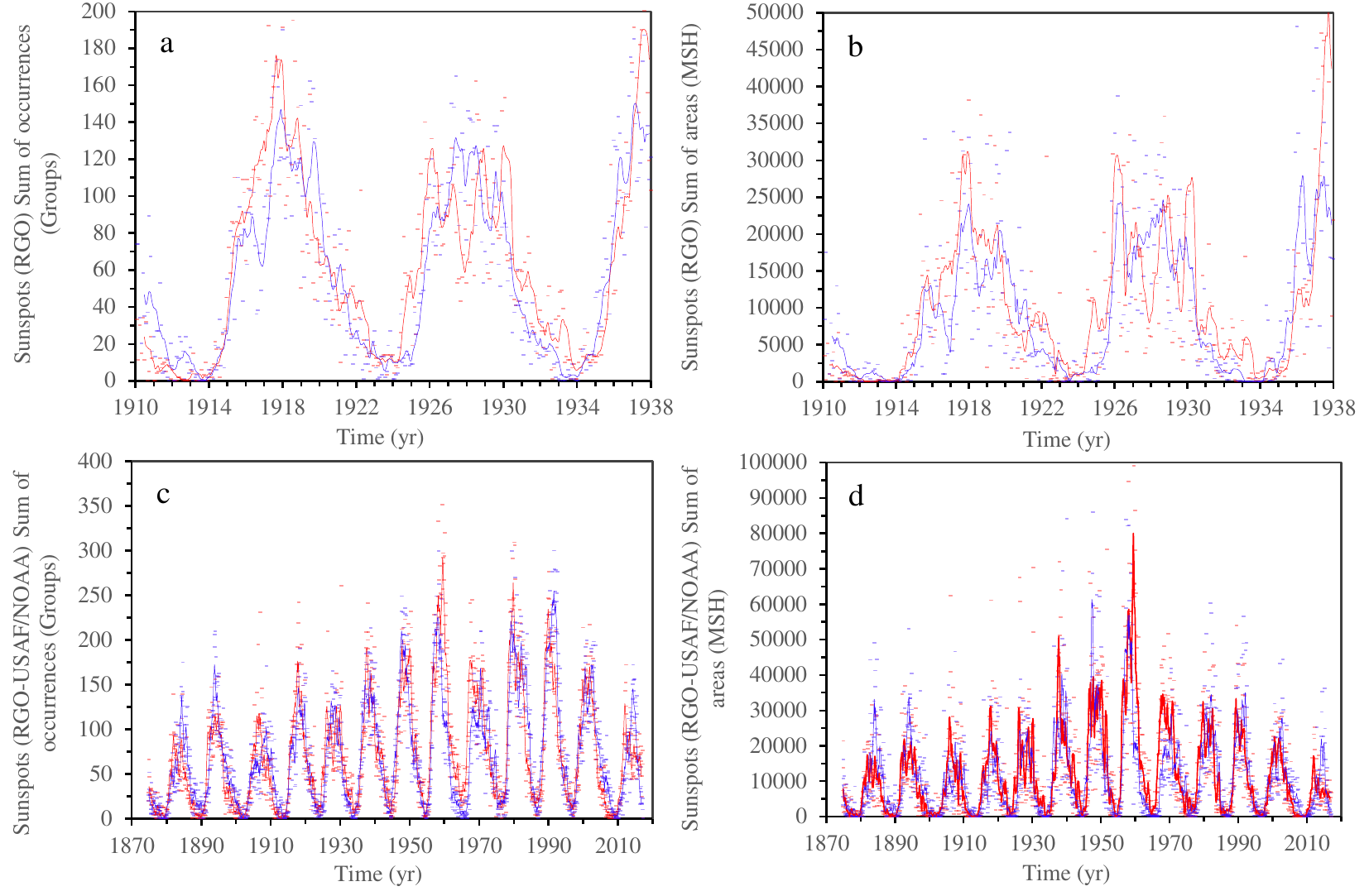}
    \caption{Evolution over time of the sunspot groups monthly sum of occurrences (left panels) and sum of areas (right panels) in the periods 1910—1937 (upper panels) and 1874—2016 (lower panels). All the data correspond to the RGO-USAF/NOAA. Pale red and blue dots represent respectively, the monthly data corresponding to the northern and southern hemispheres. Red and blue solid lines are related to the respective 7-month running averages.}
    \label{fig:Figure2}
\end{figure*}

Figure \ref{fig:Figure2} shows a comparison between both solar hemispheres in the monthly number of sunspot groups occurrences recorded by the RGO-USAF/NOAA (left panels), and their sum of areas expressed in MSH (right panels) for the periods 1910--1937 (upper panels) and 1874--2016 (lower panels). As regards the period 1910--1937, it can be seen that the RGO-USAF/NOAA numbers in both hemispheres are about $38 \pm 4 \%$ higher in comparison to those obtained for EO. This is due to the higher temporal coverage within the considered period. However, similar results as regards the monthly sum of areas and occurrences of sunspot groups and the integral of each quantity over the period considered were obtained. Thus, from RGO records, the northern hemisphere presents, on average, values which are about $5.1 \pm 0.5 \%$ higher in comparison with the southern, and also dominates the southern during the $53.9 \pm 0.4 \%$ of the total considered time-span.

\section{Analysis and results}

Section~\ref{sec:N-S asymmetry indices analysis} presents the cyclic behaviour in the N-S asymmetry of all the solar structures and properties considered as we originally figured out in the examination of the N-S asymmetry indices during 1910--1937. Once the periodic behaviour was detected, we did a thorough characterisation of the phenomenon by using different methods and techniques, i.e., the power spectrum (DFT, TAWS, and LSP), CEEMDAN and wavelet (Morlet) analyses, as Sections \ref{sec:Power spectrum analysis}, \ref{sec:CEEMDAN analysis} and \ref{sec:Wavelet analysis} show, which lead us to the detection of additional periodic signals. It is important to remark that all uncertainties in this article are computed as the standard error of the mean.

\subsection{N-S asymmetry indices analysis}
\label{sec:N-S asymmetry indices analysis}

\begin{figure*}
	\includegraphics[width=\textwidth]{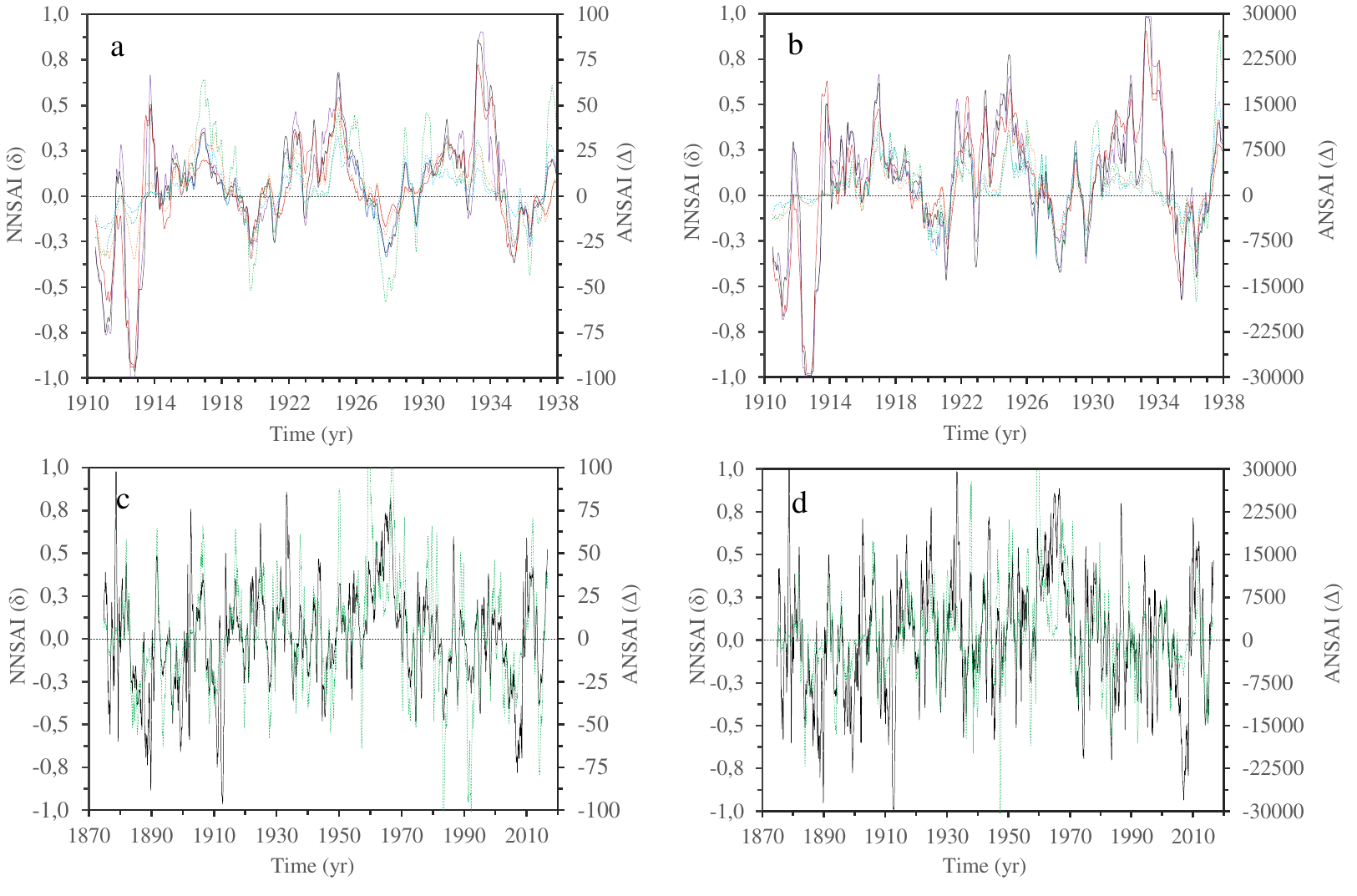}
    \caption{Evolution over time of the NNSAI and ANSAI in the periods 1910—1937 (upper panels) and 1874—2016 (only RGO-USAF/NOAA data, lower panels) of the monthly sum of occurrences (left panels) and sum of areas (right panels) of sunspots and solar plage groups. Regarding the NNSAI: Violet solid line corresponds to EO sunspots groups records, black solid line belongs to RGO-USAF/NOAA sunspots groups records, and red solid line is related to EO solar plage records. Regarding the ANSAI: Blue dashed line corresponds to EO sunspots groups records, green dashed line belongs to RGO-USAF/NOAA sunspots groups records, and orange dashed line is related to EO solar plage records.}
    \label{fig:Figure3}
\end{figure*}

Figure \ref{fig:Figure3} shows the absolute and normalised N-S asymmetry indices (hereafter ANSAI and NNSAI, respectively) in the monthly sum of sunspot and solar plage groups occurrences (left panels) and areas (right panels), recorded by the EO and by the RGO in the periods 1910--1937 (upper panels) and 1874--2016 (only RGO-USAF/NOAA data, lower panels). All the ANSAI ($\Delta$) have been calculated by using Expression~(\ref{eq:ANSAI}), where $A_N$ and $A_S$ represent respectively, the solar activity in the northern and southern hemispheres, in terms of the total number of occurrences or the sum of areas per month. Analogously, all the NNSAI ($\delta$) have been computed by using Equation~(\ref{eq:NNSAI}).

\begin{equation}
    \Delta=A_N-A_S
	\label{eq:ANSAI}
\end{equation}

\begin{equation}
    \delta=\frac{A_N-A_S}{A_S+A_S}
	\label{eq:NNSAI}
\end{equation}

In panels \textit{a} and\textit{ b}, it can be observed that solar activity is not only asymmetric for both hemispheres during all the considered period, but also slightly different in relation to the different layers of the solar atmosphere. Actually, as shown in Table~\ref{Table1}, the Pearson’s correlation coefficients calculated between the different asymmetry indices of sunspot and solar plage groups reflect these differences. Moreover, in comparison with sunspots, solar plages tend to present lower values of asymmetry in both properties analysed in practically all the studied period. Actually, in the particular case of all the NNSAI series related to the monthly sum of areas and occurrences of both solar structures, the integral over the period 1910--1937 of all data points in absolute value  is, on average, about $27.4\pm0.6\%$ lower for plages in comparison to sunspots.  

\begin{table}
\centering
\caption{Pearson’s correlation coefficient computed between the different ANSAI and NNSAI related to the monthly sum of occurrences and areas of sunspot and solar plage groups.}
\label{Table1}
\begin{tabular}{p{4.9cm}p{1.2cm}p{1.2cm}}
\hline
& \textbf{$r_{NNSAI}$} & \textbf{$r_{ANSAI}$} \\ \hline
Sunspots OE - Sunspots RGO (Occurrences) & 0.953             & 0.967             \\
Sunspots OE - Plages OE (Occurrences)    & 0.872             & 0.774             \\
Plages OE - Sunspots RGO (Occurrences)   & 0.903             & 0.735             \\
Sunspots OE - Sunspots RGO (Areas)       & 0.979             & 0.925             \\
Sunspots OE - Plages OE (Areas)          & 0.918             & 0.847             \\
Plages OE - Sunspots RGO (Areas)         & 0.904             & 0.816             \\ \hline
\end{tabular}
\end{table}

Furthermore, in both structures, a cyclic change in the predominant hemisphere can also be observed every $7.9 \pm 0.2~yr$, on average. More specifically, this phenomenon has associated a global period of approximately $7.9 \pm 0.3~ yr$ in the case of sunspot groups and $8.0 \pm 0.4 ~yr$ in the case of solar plage groups. In addition, as seen in Sections \ref{sec:EOdata} and \ref{sec:RGOdata}, the northern hemisphere presents slightly higher values of asymmetry in all sets of data, and also dominates for longer periods of time, in comparison with the southern hemisphere \citep{dePaula2020}. Concretely, during 1910-1937, all the integrals of the absolute values of each asymmetry indices are, on average, about $33 \pm 3 \%$ higher in comparison to those related to the southern hemisphere.  

\subsection{Power spectrum analysis}
\label{sec:Power spectrum analysis}

In order to confirm the initial result showed in Section~\ref{sec:N-S asymmetry indices analysis}, to obtain more accurate information about the length of the $7.9 \pm 0.2~yr$ periodicity found during 1910--1937 and, even to detect additional periodic signals, we carried out an analysis of the power spectrum of all the N-S asymmetry time series by using three different techniques: the Discrete Fourier Transform, the Lomb-Scargle Periodogram, and the Time Averaged Wavelet Spectrum. Each of the three methods is good in some aspects and not so much in others, so that all results obtained are complementary among them and allow a better understanding of the phenomenon in all its complexity.

We first applied the classical method of the Discrete Fourier Transform (DFT) to all the ANSAI related to the mentioned properties. The reason why only the ANSAI are analysed is because in contrast to them, the denominator of the NNSAI (Equation~(\ref{eq:NNSAI})) introduces additional spurious peaks, as demonstrated by \citet {Ballester2005} from the properties of DFTs. We set the following parameters to each series analysed with this method: one data per month, a sampling interval of $\Delta t = 1/12~yr$, and a Nyquist frequency of $\nu_{Ny} = 1/(2\Delta t)=6~yr^{-1}$. 

For its part, the Lomb-Scargle periodogram \citep{Lomb1976, Scargle1982} is a powerful tool commonly used in astrophysics to search for periodic patterns in an unevenly-sampled time series (see \citet {Vanderplas2018} for the Python implementation used in this Section). The LSP has already been used to investigate periodicities in the sunspot areas/numbers \citep {Ballester2005, Lomb_2013,Chowdhury_2013,Chowdhury2015,Lopes2015,Zhu2018} and also in the solar plage areas \citep {Chowdhury2016}. Actually, in this work, each obtained dataset of monthly sum of areas and occurrences of sunspots or solar plages is an uneven-time series due to the days without observations. An advantage of this method is that allows to fix the number of frequencies with which the power spectrum is evaluated (which in the DFT is limited to $n=N/2$, where $N$ is the total number of data points of a particular time series). Consequently, a more accurate estimation of the period can be achieved. Concretely, by using the LSP on the 1910—1937 series we increased the number of frequencies from 168 $(N=336)$ to 10000, evenly distributed between $100~yr^{-1}$ and the Nyquist frequency. The choice of this value has been made according to \citet {Vanderplas2018} , who argued that the minimum number of frequencies for which a LSP should be evaluated, $N_{eval}$, must be large enough to properly sample all power peaks, and is given by Equation (44) of \citet {Vanderplas2018}, which in our particular case reduces to $N_{eval}=n_o N/2$, where $n_o$ is the number of frequencies per peak and is usually taken between 5 and 10. Consequently, in the particular case of the 1910-1937 series, $N_{eval}$ is required to be, at least, about 840-1680, depending on the $n_o$ value. Thus, the choice of $N_{eval}=10000$ is more than enough to produce a well sampled LSP.

Furthermore, we also used the Python implementation of \citet{Torrence1998} for the computation of the Time Averaged Wavelet Spectrum (TAWS) (\url{https://github.com/chris-torrence/wavelets}), called originally Global Wavelet Spectrum (GWS) in \citet{Torrence1998}. The TAWS is a power spectrum whose peaks are much smoother in comparison with the two above-mentioned methods \citep{Hudgins1993,Torrence1998}.This feature can be useful to detect periodicities with a time-varying period. The TAWS can be derived from the Continuous Wavelet Transform, $CWT_n(s)$ (see Equation~(\ref{eq:CWT})), which is defined as the convolution of the discrete time series to analyse and the complex conjugate of a scaled and translated version of a certain mother function, $\psi_0(\eta)$, which can be parametrised in terms of its temporal extension and frequency. 

\begin{equation}
    CWT_n(s)=\sum_{n'=0}^{N-1}x_{n'}\psi_0^*\left[\frac{(n'-n)\Delta t}{s}\right]
	\label{eq:CWT}
\end{equation}

In our work, we used the Morlet wavelet function, defined as the product of a complex
exponential wave and a Gaussian envelope:

\begin{equation}
    \psi_0(\eta)=\frac{1}{\sqrt[4]{\pi }}e^{-\frac{\eta^2}{2}}e^{j\omega_0\eta}
    \label{Psi0_of_eta}
\end{equation}

where $j\equiv\sqrt{-1}$ denotes the imaginary unit, $\eta$ is the non-dimensional time, and $\omega_0$ is the non-dimensional central frequency of the mother function, and must be equal or higher than $6$ to satisfy the admissibility condition \citep{Farge1992}.

Finally, the TAWS is obtained by using Equation (\ref{eq:TAWS}):

\begin{equation}
    TAWS(s)=\frac{1}{N}\sum_{n'=0}^{N-1}|CWT_n(s)|^2
	\label{eq:TAWS}
\end{equation}

It is necessary to choose a set of scales $s$ in which to evaluate the TAWS. According to \citet{Torrence1998}, if the Morlet mother function is used, it is recommended to use an exponential discretization:

\begin{equation}
    s_l=s_02^{l\delta_l}, l=0,...,L
	\label{eq:sl}
\end{equation}

where $s_0=2\Delta t$, $L=\frac{1}{\delta_l}\log_2\left(\frac{N\Delta t}{s_0}\right)$, and $\delta l$ must be small enough to generate a well sampled power spectrum, i.e., equal to or lower than 0.5 \citep{Torrence1998}.In this work, we fixed $\delta_l=10^{-2}$ to achieve a higher resolution (which gives $n=740$ frequencies in the particular case of the 1910-1937 time series). Moreover, it is important to remark that the scale $s$ in which the peaks of TAWS occurs is related to their corresponding frequency (period). This relationship is given by:

\begin{equation}
    \nu=\frac{1}{s\Delta t \lambda}
	\label{eq:nu-scale}
\end{equation}

where $\lambda=\frac{4\pi s}{\omega_0 + \sqrt{2+\omega_0^2}}$ is the Fourier wavelength (frequency Fourier factor) for the particular case of the Morlet mother function \citep{Torrence1998}.

The confidence levels in DFT, TAWS, and LSP were obtained respectively by following the procedures in Sections 4.1 and 4.2 in \citet{Auchere2016}, and in Section 7.4 in \citet {Vanderplas2018}. It must be pointed that although the three procedures are similar, there are some differences among them, since each method of analysis uses a different approach (see Table~\ref{methods}).

\begin{table*}
\centering
\caption{Brief summary of the different procedures followed in each method of power spectrum analysis.}
\label{methods}
\begin{tabular}{lp{16cm}}
\hline
\textbf{Method} &
\textbf{Procedure}
  \\ \hline
DFT &
  After setting a global confidence level, $P_g$, we computed the minimum power $pm$ that peaks of the normalised DFT must present to be considered statistically significant. To do this, we used $pm=-ln\left[1-\left(1-P_g\right)^{2/N}\right]$ \citep{Auchere2016}. Finally, we extrapolated the result to the original DFT.  
  \\
  \\
TAWS &
  After setting a global confidence level, $P_g$, we computed the minimum power $pm$ that peaks of the normalised TAWS must present to be considered statistically significant. To do this, we used and inverted $P_g=1-\{1-[P(pm)]^r\}^s$, where $P(pm)$ is the local confidence level (and is computed by using the incomplete gamma function), and $r$ and $s$ are parameters empirically calculated in \citet{Auchere2016}; these authors represent them with the letters $a$ and $n$, respectively. Finally, we extrapolated the result to the original TAWS.  
  \\
  \\
LSP &
  After setting a global confidence level, $P_g$, we determined the false alarm probability, $FAP=1-P_g$, of the normalized LSP. The FAP measures the probability that a peak of a certain power could be due to Gaussian noise. Subsequently, we computed $pm$ by using the \textit{bootstrap} algorithm \citep {Vanderplas2018}. Finally, we extrapolated the result to the original LSP.  
  \\ 
\hline
\end{tabular}
\end{table*}

Only it is possible to compute the confidence levels of a power spectra if it is flat, i.e., it does not show any trend. Since the considered time series do not lead to flat mean power spectra, we normalised them by dividing them by their corresponding estimation of the Fourier background power spectrum (FBS), that is, the corresponding DFT trend in terms of the frequency $\nu$ \citep{Auchere2016}. To obtain each FBS, we fitted functions with the form: 

\begin{equation}
    FBS(\nu)=a\nu^b + c
	\label{eq:FBS}
\end{equation}

where \textit{a}, \textit{b} and \textit{c} are the parameters of the model, properly chosen by two different Python libraries \textit{(scipy.optimize.curve\_fit} and \textit{lmfit)}. Then, two different fits were obtained for each time series (S for \textit{Scipy}, and L for \textit{Lmfit}). It is important to remark that FBS fits were estimated from $TAWS(s)$. This choice is because in general, TAWS are much smoother and, consequently, they do not show drastic changes in power, as occurrs with DFT or LSP \citep{Torrence1998, Auchere2016}. However, TAWS show variations on timescales much smaller than the characteristic times of FBS. For this reason, a weight was associated to each point of the TAWS curves, so that it is greater for high power values. As explained in detail in lines 93-100 of the program $confidence\_levels\_demo.pro$ \citep{Auchere2016}, available at \url{https://idoc.ias.u-psud.fr/MEDOC/wavelets_tc98}, weights are chosen to be equal to the TAWS squared as a function of frequency. 

Finally, for each time series, we chose those FBS fits that showed for a certain value of $\omega_0$ a reduced chi-squared value, $\chi^2_{df}$, closer to one \citep{Leo1992}. Each $\chi^2_{df}$ is obtained by dividing the corresponding statistic $\chi^2$ associated to the DFT derived from a certain time series, by its number of degrees of freedom, $df=n-m$, where $n$ and $m$ represent respectively, the total number of frequencies in which each DFT is evaluated $(n=N/2)$, and the number of the FBS fit parameters $(m=3)$. Moreover, in order to determine the goodness of the chosen FBS fits, we computed the probability between 0 and 1 of obtaining a value of $\chi_{df}^2$ larger or lower than a certain value, $\chi_{{df}_i}^2$, given by the different chosen FBS fits. We considered reliable the results related to a given FBS fit as long as the probabilities $Pr(\chi_{df}^2\leq \chi_{{df}_i}^2)<0.99$ for $\chi_{{df}_i}^2>1$, and $Pr(\chi_{df}^2> \chi_{{df}_i}^2)<0.99$ for $\chi_{{df}_i}^2\leq 1$.

\begin{figure*}
	\includegraphics[width=\textwidth]{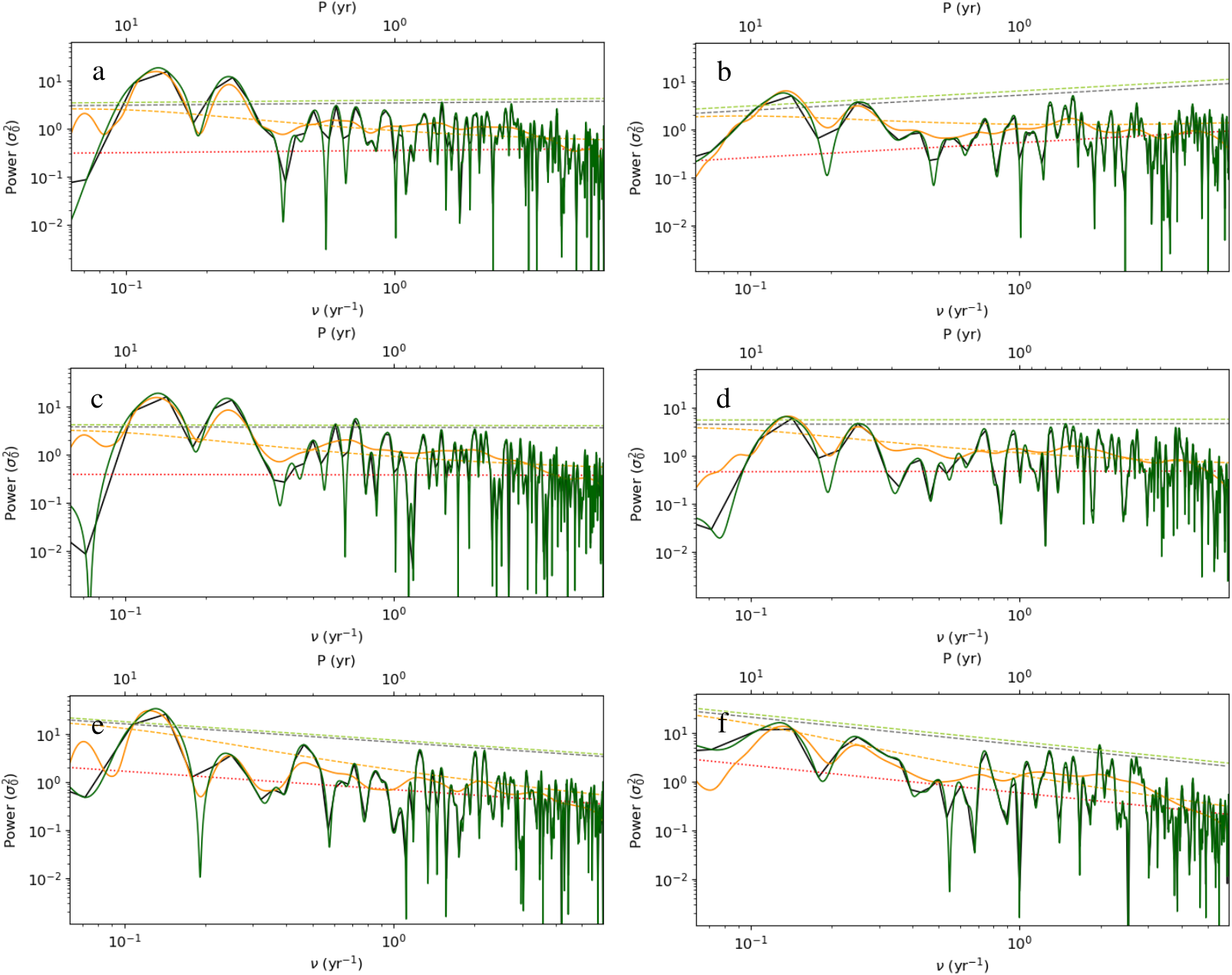}
    \caption{DFT, TAWS and LSP for all the ANSAI related to the monthly sum of occurrences (left panels) and areas (right panels) of sunspot and solar plages groups during 1910—1937. The 2-set of panels \textit{a-b}, and \textit{c-d} are related respectively to sunspot data from the EO and RGO-USAF/NOAA. Panels \textit{e-f} are related to solar plage data from EO. The continuous black, orange and green lines are the DFT, TAWS and LSP, while the dotted lines of the same colour provide their global confidence level. The dotted red line is the fitted FBS. All the confidence levels are fixed at $99\%$.}
    \label{fig:Figure4}
\end{figure*}

All the obtained results during 1910--1937 are shown in Tables \ref{Table2}, \ref{Table3} and \ref{Table4} and also plotted in Figure~\ref{fig:Figure4}.\footnote{The FBS model of Equation~(\ref{eq:FBS}) does not perfectly capture all trends of the spectra. This is the case of the power decrease at large frequencies present in panels a, c, and d in Figure~\ref{fig:Figure4}. We have tried some variations of Equation~(\ref{eq:FBS}), like adding another power function, but the fit to the spectrum yields a negligible amplitude for one of the two power functions. Then, we are left with one power function, as in Equation~(\ref{eq:FBS}). We are aware that the fits to the spectra may not be the best ones and this implies that at large frequencies the confidence levels obtained in this work can be over or underestimated.} Thus, in general, two statistically significant periodicities between 7.0-7.9 yr and 4.0-4.2 yr can be inferred. The first periodicity presents a 7.00 yr period in all cases by using the DFT, and an averaged value of $7.59\pm0.09~ yr$ and $7.57\pm0.03~yr$ by analysing all the features and solar structures via TAWS and LSP, respectively. In the case of the second periodicity, it presents a 4.00 yr period in all cases by using the DFT, and an averaged value of $4.05\pm0.02~yr$ and $4.10\pm0.04~yr$ via TAWS and LSP, respectively. It has to be mentioned that this second periodic signal is not found to be statistically significant for solar plage occurrences.

Furthermore, the TAWS and LSP analyses of both periodicities reveal that in comparison with sunspots, solar plages groups present slightly larger values of the period in both areas and occurrences. More precisely, by using the LSP we obtained a value of 7.53±0.03 yr and 4.07±0.05 yr for sunspot groups and 7.65±0.01 yr and 4.14±0.08 yr for solar plages.

\begin{table}
\centering
\caption{Non-dimensional central frequency and reduced chi-squared statistic values associated to each fit of the FBS for all the ANSAI related to the monthly sum of occurrences and areas of all sunspot and solar plages groups series during 1910--1937. Values in bold type indicate the best (and chosen) fit. Fifth column expresses the corresponding probabilities $Pr(\chi_{df}^2\leq \chi_{{df}_i}^2)$ for those chosen fits with $\chi_{{df}_i}^2>1$, or the probabilities $Pr(\chi_{df}^2> \chi_{{df}_i}^2)$ for those chosen fits with $\chi_{{df}_i}^2\leq 1$. The probability for a hypothetical fit with $\chi_{{df}_i}^2=1$ is of 0.51.}

\label{Table2}
\begin{tabular}{lllll}
\hline
\textbf{Feature (Observatory)} &
\textbf{$\omega_0$}&
\textbf{$\chi_{{df}_i}^2$~(S)} & \textbf{$\chi_{{df}_i}^2$~(L)} & \textbf{Pr} \\ \hline
Sunspot areas (OE) & 
$6.6$ & 0.8135 & \textbf{0.8341} & 0.94\\
Sunspot areas (RGO)          & 
$6.7$ & 1.6703 & \textbf{1.2523} & 0.98\\
Plage areas (OE)       & 
$6.0$ & 7.1856 & \textbf{1.0557} & 0.70\\
Sunspot occurrences (OE)     & 
$7.9$ & 1.8050 & \textbf{0.9424} & 0.69\\
Sunspot occurrences (RGO)    & 
$6.6$ & 2.1129 & \textbf{0.9985} & 0.49\\
Plage occurrences (OE) & 
$7.4$ & 2.4420 & \textbf{1.0005} & 0.52\\ \hline
\end{tabular}
\end{table}

\begin{table*}
\centering
\caption{Length of the periods obtained for the first periodic signal via DFT, TAWS and LSP for all the ANSAI related to the monthly sum of occurrences and areas of all sunspot and solar plages groups series during 1910--1937. Values in the right parentheses indicate in percentage the confidence level of the signals.}
\label{Table3}
\begin{tabular}{llll}
\hline
\textbf{Feature (Observatory)}              & 
\textbf{Period DFT}                         & 
\textbf{Period TAWS}                        & 
\textbf{Period LSP}                         \\ \hline
Sunspot areas (OE)                          & 
$7.00 ~yr (\nu=0.1428 ~yr^{-1})   (99$\%$)$ & 
$7.40 ~yr (\nu=0.1351 ~yr^{-1})   (99$\%$)$ & 
$7.56 ~yr (\nu=0.1322 ~yr^{-1})   (99$\%$)$ \\
Sunspot areas (RGO)                         & 
$7.00 ~yr (\nu=0.1428 ~yr^{-1})   (99$\%$)$ & 
$7.24 ~yr (\nu=0.1380 ~yr^{-1})   (99$\%$)$ & 
$7.42 ~yr (\nu=0.1346 ~yr^{-1})   (99$\%$)$ \\
Solar plage areas (OE)                      & 
$7.00 ~yr (\nu=0.1428 ~yr^{-1})  (<90\%)$   & 
$7.48 ~yr (\nu=0.1338 ~yr^{-1})  (99$\%$)$  & 
$ 7.70 ~yr (\nu=0.1298 ~yr^{-1}) (90$\%$)$ \\
Sunspot occurrences (OE)                    & 
$7.00 ~yr (\nu=0.1428 ~yr^{-1})   (99$\%$)$ & 
$7.75 ~yr (\nu=0.1291 ~yr^{-1})   (99$\%$)$ & 
$7.60 ~yr (\nu=0.1316 ~yr^{-1})   (99$\%$)$ \\
Sunspot occurrences (RGO)                   & 
$7.00 ~yr (\nu=0.1428 ~yr^{-1})   (99$\%$)$ & 
$7.61 ~yr (\nu=0.1314 ~yr^{-1})   (99$\%$)$ & 
$7.53 ~yr (\nu=0.1328 ~yr^{-1})   (99$\%$)$ \\
Solar plage occurrences (OE)                & 
$7.00 ~yr (\nu=0.1428 ~yr^{-1})   (99$\%$)$ & 
$7.92 ~yr (\nu=0.1262 ~yr^{-1})   (99$\%$)$ & 
$7.60 ~yr (\nu=0.1316 ~yr^{-1})   (99$\%$)$ \\ \hline
\end{tabular}
\end{table*}

\begin{table*}
\centering
\caption{Same as Table~\ref{Table3}, but for the second periodic signal during 1910--1937.}
\label{Table4}
\begin{tabular}{llll}
\hline
\textbf{Feature (Observatory)}           & 
\textbf{Period DFT}                      & 
\textbf{Period TAWS}                     & 
\textbf{Period LSP}                      \\ \hline
Sunspot areas (OE)                       & 
$4.00 yr (\nu=0.2500 ~yr^{-1})   (99\%)$ &
$3.99 yr (\nu=0.2504 ~yr^{-1})   (99\%)$ & 
$3.96 yr (\nu=0.2526 ~yr^{-1})   (95\%)$ \\
Sunspot areas (RGO)                      & 
$4.00 yr (\nu=0.2500 ~yr^{-1})   (99\%)$ & 
$3.99 yr (\nu=0.2506 ~yr^{-1})   (99\%)$ & 
$3.99 yr (\nu=0.2508 ~yr^{-1})   (95\%)$ \\
Solar plage areas (OE)                   &
$4.00 yr (\nu=0.2500 ~yr^{-1}) (<90\%)$   &
$4.01 yr (\nu=0.2496 ~yr^{-1}) (99\%)$   &
$4.08 yr (\nu=0.2448 ~yr^{-1}) (<90\%)$   \\
Sunspot occurrences (OE)                 & 
$4.00 yr (\nu=0.2500 ~yr^{-1})(99\%)$    & 
$4.12 yr (\nu=0.2426 ~yr^{-1})   (99\%)$ & 
$4.16 yr (\nu=0.2406 ~yr^{-1})   (99\%)$ \\
Sunspot occurrences (RGO)                & 
$4.00 yr (\nu=0.2500 ~yr^{-1})   (99\%)$ & 
$4.14 yr (\nu=0.2418 ~yr^{-1})   (99\%)$ & 
$4.18 yr (\nu=0.2394 ~yr^{-1})   (99\%)$ \\
Solar plage occurrences (OE)             &
$4.00 yr (\nu=0.2500 ~yr^{-1}) (<90\%)$  &
$4.05 yr (\nu=0.2472 ~yr^{-1}) (<90\%)$   &
$4.20 yr (\nu=0.2382 ~yr^{-1}) (<90\%)$   \\ \hline
\end{tabular}
\end{table*}

\begin{figure*}
	\includegraphics[width=\textwidth]{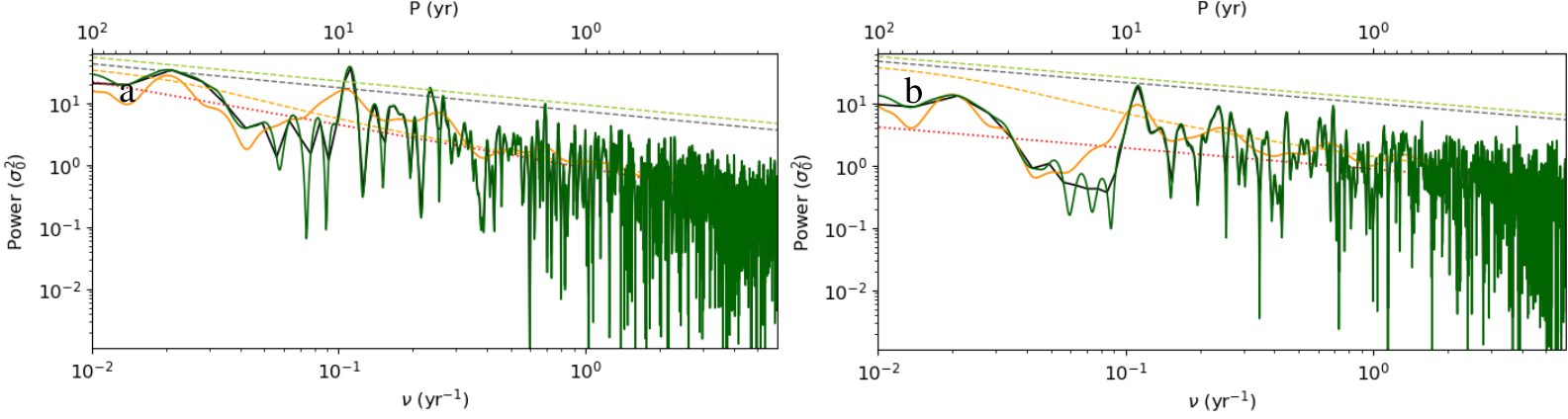}
    \caption{DFT, TAWS and LSP for the ANSAI related to the monthly sum of occurrences (panel \textit{a}) and areas (panel \textit{b}) of sunspot groups registered by the RGO-USAF/NOAA during 1874—2016. The continuous black, orange and green lines are the DFT, TAWS and LSP, while the dotted lines of the same colour provide their global confidence level. The dotted red line is the fitted FBS. All the confidence levels are fixed at $99\%$.}
    \label{fig:Figure5}
\end{figure*}

For the purpose of checking whether the cyclic behaviour observed during 1910--1937 is maintained over the time, we did the same spectral analysis by using the full series of sunspot data recorded by the RGO-USAF/NOAA during 1874--2016. In this case, N=1710 so $N_{eval}$ is required to be, at least, about 4275-8550, depending on the $n_0$ value. We decided to maintain $N_{eval}=10000$. Table~\ref{Table5} shows the $\chi_{{df}_i}^2$ of the different fitted FBS. Taking advantage of the TAWS properties, in both areas and occurrences data series it is possible to distinguish three statistically significant periodicities of $1.48\pm0.01~yr$, $3.98\pm0.03~yr$, and $9.1\pm0.1~yr$, as indicated in detail in Table~\ref{Table6} and shown in Figure~\ref{fig:Figure5}. In the DFT and LSP curves, these peaks are clearly visible, but they present a slightly lower significance. Moreover, it is also remarkable the unnoticed peak around $7.19\pm0.03~yr$ that can be observed in the DFT and LSP curves, but it seems to be diluted in the TAWS. This fact may due to phenomena as the temporal drift and intermittence, which weakens the periodic signal.

\begin{table}
\centering
\caption{Non-dimensional central frequency and reduced chi-squared statistic values associated to each fit of the FBS for the ANSAI related to the monthly sum of occurrences and areas of the entire RGO-USAF/NOAA sunspot groups series. Values in bold type indicate the best (and chosen) fit. Fifth column expresses the corresponding probabilities $Pr(\chi_{df}^2\leq \chi_{{df}_i}^2)$ for that chosen fit with $\chi_{{df}_i}^2>1$, or the probabilities $Pr(\chi_{df}^2> \chi_{{df}_i}^2)$ for that chosen fit with $\chi_{{df}_i}^2\leq 1$. The probability for a hypothetical fit with $\chi_{{df}_i}^2=1$ is of 0.51.}

\label{Table5}
\begin{tabular}{lllll}
\hline
\textbf{Feature} &
\textbf{$\omega_0$} &
\textbf{$\chi_{{df}_i}^{2}$~(S)} & \textbf{$\chi_{{df}_i}^{2}$~(L)}  & \textbf{Pr}\\ \hline
Areas            & 
$7.5$ & 1.0002 & \textbf{1.0002}   & 0.51 \\
Occurrences      & 
$7.5$ & \textbf{0.9977} & 0.9063   & 0.51 \\ \hline
\end{tabular}
\end{table}

\begin{table*}
\centering
\caption{Length of the periods obtained for all the detected periodicities via DFT, TAWS and LSP for the ANSAI related to the monthly sum of occurrences and areas of the entire RGO-USAF/NOAA sunspot groups series. Values in the right parentheses indicate in percentage the confidence level of each periodic signal. Dashed values indicate the absence of the period}
\label{Table6}
\begin{tabular}{llll}
\hline
\textbf{Feature}                         & 
\textbf{Period DFT}                      & 
\textbf{Period TAWS}                     & 
\textbf{Period LSP}                      \\ \hline
Areas                                    &  
$8.90 yr (\nu=0.1123 ~yr^{-1})(95\%)$            & 
$8.92 yr (\nu=0.1121 ~yr^{-1})(99\%)$            & 
$8.90 yr (\nu=0.1124 ~yr^{-1})(<90\%)$ 
\\& 
$7.13 yr (\nu=0.1403~yr^{-1})(<90\%)$  & \multicolumn{1}{c}{-}                  & 
$7.24 yr (\nu=0.1382 ~yr^{-1})(<90\%)$ 
\\& 
$4.19 yr (\nu=0.2386 ~yr^{-1})(<90\%)$ & 
$4.02 yr (\nu=0.2488 ~yr^{-1})(99\%)$  & 
$4.21 yr (\nu=0.2376 ~yr^{-1})(<90\%)$ 
\\& 
$1.45 yr (\nu=0.6877 ~yr^{-1})(<90\%)$ & 
$1.49 yr (\nu=0.6703 ~yr^{-1})(99\%)$  & 
$1.45 yr (\nu=0.6911 ~yr^{-1})(<90\%)$ 
\\ \hline
Occurrences                            & 
$8.90 yr (\nu=0.1123 ~yr^{-1})(99\%)$  & 
$9.30 yr (\nu=0.1075 ~yr^{-1})(99\%)$  & 
$8.99 yr (\nu=0.1112 ~yr^{-1})(99\%)$ 
\\& 
$7.13 yr (\nu=0.1403 ~yr^{-1})(<90\%)$ & \multicolumn{1}{c}{-}                  & 
$7.14 yr (\nu=0.1400 ~yr^{-1})(<90\%)$ 
\\& 
$4.19 yr (\nu=0.2386 ~yr^{-1})(99\%)$ & 
$3.94 yr (\nu=0.2540 ~yr^{-1})(99\%)$ & 
$4.24 yr (\nu=0.2358 ~yr^{-1})(99\%)$ 
\\& 
$1.45 yr (\nu=0.6877 ~yr^{-1})(99\%)$ & 
$1.47 yr (\nu=0.6797 ~yr^{-1})(99\%)$ & 
$1.45 yr (\nu=0.6893 ~yr^{-1})(95\%)$ 
\\ \hline
\end{tabular}
\end{table*}

We also carried out the same calculation by dividing the 1814--2016 RGO-USAF/NOAA data in twelve intervals of time of three solar cycles each, by taking the $(i-1)-th$, $i-th$, and $(i+1)-th$ solar cycles, and considering $i$ from the 12th to the 23rd Solar Cycles. All the information concerning the $\chi_{{df}_i}^{2}$ of the different FBS proposed for all the intervals can be found in Table~\ref{Table7}, and the evolution over time of the five periodicities detected in the occurrences and areas data, is disposed in Tables ~\ref{TableS-1} and ~\ref{TableS-2}. Due to the length of these two last tables, only the results related to the 11-12-13 Solar Cycles are displayed, which correspond to years 1874--1902. The full tables are available online as a supplementary material (see Tables S-1 and S-2). It has to be mentioned that by following the above-mentioned criterion of reliability, results derived from occurrences data belonging to Solar Cycles 16-17-18, 17-18-19, and 22-23-24 are not reliable enough. 

\begin{table*}
\centering
\caption{Non-dimensional central frequency and reduced chi-squared statistic values associated to each fit of the FBS for all the ANSAI related to the monthly sum of areas and occurrences of the RGO-USAF/NOAA sunspot groups series divided in twelve intervals of time in sets of three solar cycles. Values in bold type indicate the best (and chosen) fit. Fifth and ninth columns express the corresponding probabilities $Pr(\chi_{df}^2\leq \chi_{{df}_i}^2)$ for those chosen fits with $\chi_{{df}_i}^2>1$, or the probabilities $Pr(\chi_{df}^2> \chi_{{df}_i}^2)$ for those chosen fits with $\chi_{{df}_i}^2\leq 1$. The probability for a hypothetical fit with $\chi_{{df}_i}^2=1$ is of 0.51.}
\label{Table7}
\begin{tabular}{lllll|llll}
\hline
\textbf{Solar Cycles} & 
\textbf{$\omega_0$} & 
\textbf{$\chi_{{df}_i}^{2}$~(Areas; S)} & \textbf{$\chi_{{df}_i}^{2}$~(Areas; L)} &
\textbf{Pr} &
\textbf{$\omega_0$} & 
\textbf{$\chi_{{df}_i}^{2}$~(Occurrences; S)} & \textbf{$\chi_{{df}_i}^{2}$~(Occurrences; L)} &
\textbf{Pr} \\ \hline

11-12-13 & 
6.0 & 1.9805 & \textbf{0.8533} & 0.92 & 6.0 & \textbf{0.9869} & 0.9869 & 0.53
\\

12-13-14              &        
6.0 & \textbf{1.1445} & 1.1445 & 0.93 & 8.0 & \textbf{1.1133} & 1.1133 & 0.87
\\

13-14-15              &            
6.0 & \textbf{1.0643} & 0.9202 & 0.74 & 8.0 & \textbf{1.0828} & 1.0828 & 0.80
\\

14-15-16              & 
6.0 & 1.3600 & \textbf{1.1254} & 0.88 & 7.8 & \textbf{0.8600} & 0.8600 & 0.91
\\

15-16-17              & 
6.0 & 1.0519 & \textbf{1.0519} & 0.70 & 6.6 & 2.0128 & \textbf{0.9993} & 0.51
\\

16-17-18              & 
6.0 & \textbf{0.8717} & 0.7654 & 0.89 & 8.0 & 1.3650 & \textbf{1.3649} & 0.9992
\\

17-18-19              & 
8.0 & 1.1706 & \textbf{1.1705} & 0.94 & 6.0 & 4.6423 & \textbf{1.2822} & 0.994
\\

18-19-20              & 
7.9 & \textbf{0.8352} & 0.7098 & 0.95 & 8.0 & \textbf{0.8610} & 0.7495 & 0.92
\\

19-20-21              & 
6.9 & 2.1094 & \textbf{1.0023} & 0.52 & 6.0 & 1.1900 & \textbf{0.7495} & 0.56
\\

20-21-22              & 
6.0 & 1.1203 & \textbf{1.1191} & 0.87 & 8.0 & 1.1687 & \textbf{1.1686} & 0.94
\\

21-22-23              & 
6.0 & \textbf{0.9480} & 0.9431 & 0.68 & 7.6 & \textbf{1.0002} & 0.8097 & 0.51
\\

22-23-24              & 
7.3 & 1.2387 & \textbf{1.1879} & 0.96 & 8.0 & 3.0536 & \textbf{1.6250} & 0.9999998
\\
\hline
\end{tabular}
\end{table*}

\begin{table*}
\centering
\caption{Length of the periods obtained for the five periodic signals via DFT, TAWS and LSP for all the ANSAI related to the monthly sum of areas of the sunspot groups series recorded by the RGO-USAF/NOAA during 1874—1902. Values in the right parentheses indicate in percentage the confidence level of the periodic signals. Dashed values indicate the absence of the period.This is a sample table. The full table is available online as a supplementary material and contains the results concerning years 1874--2016 (See Table S-1).}
\label{TableS-1}
\begin{tabular}{lllll}
\cline{1-4}
\multicolumn{1}{c}{\textbf{Solar Cycles}} & \multicolumn{1}{c}{\textbf{Period DFT}} & \multicolumn{1}{c}{\textbf{Period TAWS}} & \multicolumn{1}{c}{\textbf{Period LSP}} &  \\ \cline{1-4}
                                          & -                                       & -                                        & -                                       &  \\
\multicolumn{1}{c}{\textbf{11-12-13}}     & $9.44 yr (\nu=0.1059 yr^{-1})(95\%)$           & $10.79 yr (\nu=0.0927 yr^{-1})(99\%)$           & $10.52 yr (\nu=0.0951 yr^{-1})(95\%)$          &  \\
\multicolumn{1}{c}{(1874--1902)}          & $4.72 yr (\nu=0.2118 yr^{-1})(99\%)$           & $4.80 yr (\nu=0.2085 yr^{-1})(99\%)$            & $4.86 yr (\nu=0.2059 yr^{-1})(99\%)$           &  \\
                                          & $3.54 yr (\nu=0.2824 yr^{-1})(99\%)$           & $3.51 yr (\nu=0.2848 yr^{-1})(99\%)$            & $3.37 yr (\nu=0.2969 yr^{-1})(90\%)$           &  \\
                                          & $1.42 yr (\nu=0.7059 yr^{-1})(90\%)$           & $1.34 yr (\nu=0.7464 yr^{-1})(99\%)$            & $1.40 yr (\nu=0.7145 yr^{-1})(90\%)$           &  \\ \cline{1-4}
\end{tabular}
\end{table*}

\begin{table*}
\centering
\caption{Same as Table \ref{TableS-1} for the monthly sum of occurrences of the sunspot groups series recorded by the RGO-USAF/NOAA during 1874—1902. This is a sample table. The full table is available online as a supplementary material and contains the results concerning years 1874--2016 (see Table S-2).}
\label{TableS-2}
\begin{tabular}{cllll}
\cline{1-4}
\textbf{Solar Cycles} & \multicolumn{1}{c}{\textbf{Period FTT}}    & \multicolumn{1}{c}{\textbf{Period TAWS}} & \multicolumn{1}{c}{\textbf{Period LSP}}    &  \\ \cline{1-4}
                      & -                                          & -                                        & -                                          &  \\
\textbf{11-12-13}     & $9.44 yr (\nu=0.1059 yr^{-1})(99\%)$              & $10.57 yr (\nu=0.0946 yr^{-1})(99\%)$           & $10.10 yr (\nu=0.0990 yr^{-1})(99\%)$             &  \\
(1874--1902)          & $4.72 yr (\nu=0.2118 yr^{-1})(99\%)$              & $4.67 yr (\nu=0.2144 yr^{-1})(99\%)$            & $4.67 yr (\nu=0.2143 yr^{-1})(99\%)$              &  \\
\multicolumn{1}{l}{}  & $3.15 yr (\nu=0.3176 yr^{-1})(95\%)$              & $3.42 yr (\nu=0.2928 yr^{-1})(99\%)$                                             & $3.28 yr (\nu=0.3047 yr^{-1})(95\%)$              &  \\
\multicolumn{1}{l}{}  & $1.42 yr (\nu=0.7059 yr^{-1})(<90\%)$ & $1.09 yr (\nu=0.9190 yr^{-1})(95\%)$                                                            & $1.41 yr (\nu=0.7103 yr^{-1})(<90\%)$              &  \\ \cline{1-4}
\end{tabular}
\end{table*}

\begin{figure*}
	\includegraphics[width=\textwidth]{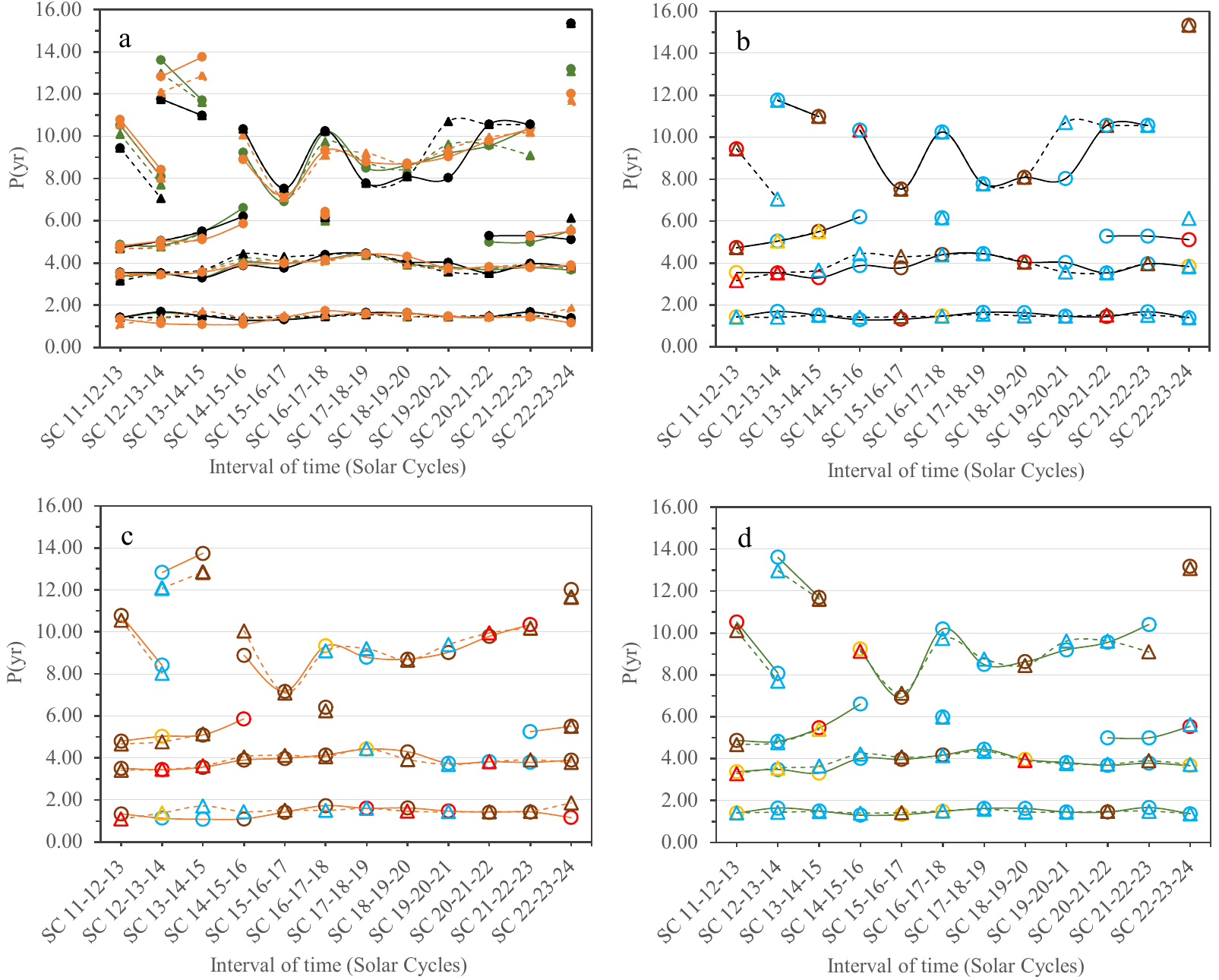}
    \caption{Evolution over the time (expressed in terms of groups of three consecutive solar cycles) of the 5 most important periodicities obtained from the ANSAI related to the monthly sum of occurrences (dashed lines and triangles) and areas (solid lines and circles) of sunspot groups data from the RGO-USAF/NOAA during 1874—2016 via DFT (black lines), TAWS (orange lines) and LSP (green lines). Panel \textit{a} shows a summary of the results without considering the confidence level associated to the signals. Panels \textit{b}, \textit{c}, and \textit{d} show the respective results obtained via DFT, TAWS and LSP taking into account 4 confidence levels ($99\%$ - brown; $95\%$- red; $90\%$ - yellow; $<90\%$ - blue) associated to the signals.}
    \label{fig:Figure6}
\end{figure*}

In order to examine the five detected periodicities, we plotted their evolution over time in Figure~\ref{fig:Figure6}, according to their confidence level $(<90\%, 90\%, 95\%$ and $99\%)$.We also computed their averaged period in Table \ref{Table10}. It can be noted that periodic signals \textit{D} and \textit{E} are quite regular in period and persist over the solar cycles. In contrast, signals \textit{A}, \textit{B}, and \textit{C} are more intermittent in time and also present important drifts on their period, which makes it difficult to follow their evolution over time and even associate each instantaneous period to a certain periodic signal. In addition, despite of the vast majority of significances are lower than $90\%$, all five periodicities can be distinguished in both, areas and occurrences data series, which make us to consider the possibility that they could not be the result of randomness, but the manifestation of relatively weak and independent periodic processes. To look at this in detail, we further studied the cyclic behaviour of the N-S asymmetry by applying the CEEMDAN and wavelet analyses to the RGO-USAF/NOAA sunspot groups data in Sections \ref{sec:CEEMDAN analysis} and \ref{sec:Wavelet analysis}, respectively. 

\begin{table}
\centering
\caption{Length of the mean periods obtained for the five periodic signals via DFT, TAWS and LSP for all the ANSAI related to the monthly sum of occurrences and areas of the sunspot groups series recorded by the RGO-USAF/NOAA during 1874--2016. }
\label{Table10}
\begin{tabular}{llll}
\hline
\textbf{Signal} & \textbf{M. Period DFT} & \textbf{M. Period TAWS} & \textbf{M. Period LSP} \\ \hline
A & 12.7 ± 0.8 yr  & 12.5 ± 0.3 yr  & 12.7 ± 0.3 yr  \\
B & 9.2 ± 0.3 yr   & 9.2 ± 0.2 yr   & 9.0 ± 0.2 yr   \\
C & 5.4 ± 0.1 yr   & 5.4 ± 0.2 yr   & 5.4 ± 0.2 yr   \\
D & 3.87 ± 0.08 yr & 3.87 ± 0.06 yr & 3.83 ± 0.06 yr \\
E & 1.48 ± 0.02 yr & 1.43 ± 0.04 yr & 1.47 ± 0.02 yr \\ \hline
\end{tabular}
\end{table}

\subsection{CEEMDAN analysis}
\label{sec:CEEMDAN analysis}

The CEEMDAN algorithm is a variation of the classical EMD (Empirical Mode Decomposition) method. Both techniques were conceived with the objective to separate or decompose a signal into oscillating components known as Intrinsic Mode Functions (IMFs). Moreover, since these techniques do not use a set of basis function, the main advantage in contrast to the power spectrum analysis is that all extracted components are anharmonic signals \citep{Kolotkov2015}, that is, they are not a product of a linear combination or a submultiple of stronger signals.

The EMD method was introduced by \citet{Huang1998} in order to deal with non-linear and non-stationary time series, and since then, it has been applied in several works related to solar processes and phenomena \citep{Terradas_2004,Qiang2007,Nakariakov2010,Barnhart_2011,Vecchio2012,Kolotkov2015, Kolotkov2016, Vecchio2017, Deng2019}. This technique is based in the decomposition of a complicated data set $x(t)$ into a finite and usually small number of $M$ $IMFs$, which represent hidden oscillation modes at different time scales. In particular, each $IMF$ presents oscillations on a larger time scale than the previous one. The algorithm also gives a residual component $r(t)$ that cannot be decomposed into $IMFs$:

\begin{equation}
    x(t)=\sum_{i=1}^M {IMF_i(t)}+r(t)
	\label{eq:x(t)}
\end{equation}

IMFs can be expressed by an amplitude $A_i(t)$ and a phase $\phi_i(t)$. They are obtained by following an iterative process called sifting, based on the identification of the relative maxima and minima from the data to be analysed \citep{Barnhart_2011,Vecchio2017}:

\begin{equation}
    IMF_i(t)=A_i(t)~cos[\phi_i(t)]
	\label{IMF_i(t)}
\end{equation}

The decomposition of the signal in $M~IMFs$ allows to define the instantaneous frequency $\nu_i(t)$ for each mode, which can be calculated by applying the Hilbert transform $\mathcal{H}$ to a certain $IMF_i(t)$ \citep{Huang1998}:

\begin{equation}
    \nu_i(t)=\frac{1}{2\pi} \frac{d~arctan[\frac{\mathcal{H}\{IMF_i(t)\}}{IMF_i(t)}]}{dt}
	\label{nu_i(t)}
\end{equation}

The major drawback of the original EMD technique is the frequent appearance of mode mixing. The mode mixing phenomenon occurs when the two or more oscillations in different time scales appear in an $IMF$, or when oscillations with a given time scale appear in different $IMFs$ \citep{Deng2019}. This issue can also be observed in crossings of the curves that represent the evolution over time of the modes’ instantaneous frequencies. To overcome the mode mixing problem, \citet{Wu2009} developed an improved method called the Ensemble Empirical Mode Decomposition (EEMD), which adds Gaussian white noise into the input signal in order to better separate the time scales. In this algorithm, the decomposition of the original signal $x(t)$ into $IMFs$ is done for $N$ realizations, adding random white noise $w_i$  each time. Thus, the general expression for the signal to treat is as follows:

\begin{equation}
    X_i(t)=x(t)+w_i ~~~(i=1,...,N)
	\label{X_i(t)}
\end{equation}

Thus, at the end of the procedure, a number of k final modes (hereafter $\overline{IMFs}$) are obtained from averaging the corresponding $IMFs$ computed via EMD in $N$ trials:

\begin{equation}
    \overline{IMF_k}(t)=\frac{1}{N}\sum_{i=1}^{N} IMF_k^i(t)  
    \label{overlineIMFk}
\end{equation}

Nevertheless, the EEMD method also has limitations since the reconstructed signal includes residual noise coming from the white noise. Moreover, different realizations of the EMD decomposition on different $X_i (t)$ can lead to a different number of $IMFs$, so the computation of each mean would be compromised. Hence, in order to overcome these inconveniences, \citet {Torres2011} proposed the CEEMDAN algorithm, which defines the treated signal as $X_i(t)$:

\begin{equation}
    X_i(t)= x(t)+\varepsilon_{i-1}w_i ~~~(i=1,...,N)
    \label{X_i(t)2}
\end{equation}

where $\varepsilon_i$ is a coefficient which allows us to select the signal-to-noise ratio at each stage, and can be fixed as a parameter $(\varepsilon_i\equiv\varepsilon)$. \citet{Wu2009} suggested to use small amplitude values for data dominated by high-frequency components, and vice versa.

In order to obtain all the final modes with this method (hereafter $\widetilde{IMFs}$), firstly the $\widetilde{IMF_1}(t)$ is computed via the EMD algorithm:

\begin{equation}
    \widetilde{IMF_1}(t)=\frac{1}{N}\sum_{i=1}^{N} IMF_1^i(t) 
    \label{widetildeIMF1}
\end{equation}

To compute the rest of $\widetilde{IMFs}$ it is necessary to define $E_j(\cdot)$ as an operator which acts on the signal producing the $j-th$ mode:

\begin{equation}
    \widetilde{IMF_k}(t)=\frac{1}{N}\sum_{i=1}^{N} E_{k-1}[r_{k-1}(t)+\varepsilon~w_i] ~~~(k=2,...,K)
    \label{widetildeIMFk}
\end{equation}

where the residues $r_k(t)$ are computed by using the following expressions:

\begin{equation}
    r_k(t) =
    \left\{
	    \begin{array}{ll}
	    	x(t)-\widetilde{IMF_1}(t)  & \mbox{if } k=1 \\
	       r_{k-1}(t)-\widetilde{IMF_{k-1}}(t) & \mbox{if } k=2,...,K
    	\end{array}
    \right.
    \label{r_k(t)}
\end{equation}

In this case, the successive $\widetilde{IMFs}$ are computed until the residue is no longer feasible to be decomposed.

The CEEMDAN Python implementation of \citet{Laszuk2017} is used here. All the results obtained by applying the CEEMDAN algorithm to the above-mentioned twelve intervals of time from the RGO-USAF/NOAA sunspot datasets during 1874--2016 are available online as three figures of eight panels each (Figures S-1, S-2, and S-3), and also are summarized in this article in Table~\ref{Table11}. Figures were obtained considering the minimum number of trials for not observing any changes between two independent runs, i.e. 15000 trials. In addition, the optimal value of $\varepsilon$ in each dataset was fixed by defining it as the one which better avoids the mode mixing problem, although, unfortunately, it is still present occasionally in some panels in the form of crossings among curves. This effect is more important in high frequency signals, since they are closer to each other. Moreover, edge effects can be noted in both ends of some modes. 

The decomposition of the input signals, i.e., the ANSAI derived from the monthly sum of areas and occurrences of sunspot groups, gives a number of $\widetilde{IMFs}$ which vary from 5 to 7 depending on the considered interval and property. Furthermore, supplementary Figures S-1, S-2, and S-3 show the evolution over time, in sets of three solar cycles, of the different instantaneous frequencies obtained from all the $\widetilde{IMFs}$ and the final residue, which gives us an idea of the stability of each oscillating mode and also the value of their period in every instant of time. For a better visualisation, all curves were smoothed by a running mean with a window width of 7 months. 

It can be seen that only the 5th oscillating mode presents, on average, an approximately stable period within each considered time range of three solar cycles. $\widetilde{IMFs}$ $\#$1 and $\#$2 present very irregular periods, and shorter than the periodic signal E detected in the previous Section; $\widetilde{IMFs}$ $\#$3 and $\#$4 present irregular periods; and $\widetilde{IMFs}$ $\#$6 and $\#$7 often present a period around 13 yr, similar to that of signal A. In some occasions, these two $\widetilde{IMFs}$ also show larger periods, whose relevance cannot be firmly established because the present analysis is applied to sets of three solar cycles, and could be related to the mode mixing phenomenon. 

In order to summarise all the results related to the $\widetilde{IMF}~\#$5, Table~\ref{Table11} presents the averaged periods for each property and set of solar cycles considered. In order to avoid edge effects in the computation of each averaged period, we did not consider the first and last years of each set of solar cycles. It can be noted in both areas and occurrences, that the 5th $\widetilde{IMF}$ does not seem an intermittent component, since it is present in all intervals of time, but does not have a completely stable instantaneous frequency over the different sets of three solar cycles. This fact suggests the presence of mode mixing among the $\widetilde{IMF}~\#$5, $\#$6 and $\#$7, since those temporal windows with a lower number of modes, present longer periods for the 5th component. Moreover, in some moments, and especially during the 15th, 16th and 17th Solar Cycles, it presents a constant value very close to 7.0-7.9 yr. This could be associated to the periodic signal we found by doing the DFT/TAWS/LSP analyses of all datasets during 1910--1937.

\begin{table*}
\centering
\caption{Length of the averaged periods obtained from averaging the $\widetilde{IMF}~\#$5 instantaneous frequency values via CEEMDAN algorithm for all the ANSAI related to the monthly sum of occurrences (O) and areas (A) of all sunspot groups recorded by the RGO-USAF/NOAA during the 19th-24th solar cycles (February 1954—October 2016). For both areas and occurrences, the peak during 1884—1891 (solar cycles 11-12-13) is not considered in the computation of the mean. The same for the occurrences, peaks during 1895—1905 (solar cycles 12-13-14) and 1896—1903 (solar cycles 13-14-15).}
\label{Table11}
\begin{tabular}{lllll}
\hline
\textbf{Solar cycles} & \textbf{Analysed years (A)} & \textbf{Analysed years (O)} & \textbf{Mean period (A)} & \textbf{Mean period (O)} \\ \hline
11-12-13 & {[}1878—1900.5{]}   & {[}1882—1898{]}     & 
6.09±0.06 yr ($\nu$=0.164±0.002   yr$^{-1}$) & 
7.6±0.2 yr ($\nu$=0.131±0.004   yr$^{-1}$)   \\
12-13-14 & {[}1882—1908{]}     & {[}1882.5—1908{]}   & 
6.5±0.1 yr ($\nu$=0.154±0.003   yr$^{-1}$)   & 
11.4±0.1 yr   ($\nu$=0.09±0.01 yr$^{-1}$)    \\
13-14-15 & {[}1892—1922{]}     & {[}1892.5—1919{]}   & 
6.41±0.05 yr ($\nu$=0.156±0.001   yr$^{-1}$) & 
9.1±0.2 yr ($\nu$=0.110±0.002   yr$^{-1}$)   \\
14-15-16 & {[}1905—1927.5{]}   & {[}1906—1930.5{]}   & 
6.0±0.1 yr ($\nu$=0.168±0.003   yr$^{-1}$)   & 
9.5±0.2 yr ($\nu$=0.105±0.002   yr$^{-1}$)   \\
15-16-17 & {[}1916—1942{]}     & {[}1915.5—1937.5{]} & 
6.47±0.05 yr ($\nu$=0.155±0.001  yr$^{-1}$) & 
7.36±0.07 yr ($\nu$=0.136±0.001   yr$^{-1}$) \\
16-17-18 & {[}1930—1952{]}     & {[}1925.5—1952{]}   & 
4.8±0.1 yr ($\nu$=0.206±0.004   yr$^{-1}$)   & 
8.8±0.1 yr ($\nu$=0.114±0.001   yr$^{-1}$)   \\
17-18-19 & {[}1934.5—1964{]}   & {[}1941—1962{]}     & 
7.3±0.1 yr ($\nu$=0.137±0.002   yr$^{-1}$)   & 
7.9±0.1 yr ($\nu$=0.127±0.002   yr$^{-1}$)   \\
18-19-20 & {[}1948.5—1971.5{]} & {[}1947—1970{]}     & 
8.52±0.06 yr ($\nu$=0.117±0.001   yr$^{-1}$) & 
8.7±0.1 yr ($\nu$=0.115±0.001   yr$^{-1}$)   \\
19-20-21 & {[}1947—1972{]}     & {[}1955.5—1984.5{]} & 
9.4±0.1 yr ($\nu$=0.106±0.001   yr$^{-1}$)   & 
10.3±0.2 yr ($\nu$=0.097±0.002   yr$^{-1}$)  \\
20-21-22 & {[}1966.5—1994{]}   & {[}1966—1996{]}     & 
9.10±0.07 yr ($\nu$=0.110±0.001   yr$^{-1}$) & 
10.3±0.2 yr ($\nu$=0.097±0.002   yr$^{-1}$)  \\
21-22-23 & {[}1880.5—1999.5{]} & {[}1978.5—2005.5{]} & 
7.3±0.1 yr ($\nu$=0.137±0.002   yr$^{-1}$)   & 
6.7±0.1 yr ($\nu$=0.149±0.002   yr$^{-1}$)   \\
22-23-24 & {[}1993—2012.5{]}   & {[}1989—2014.5{]}   & 
6.5±0.1 yr ($\nu$=0.153±0.003   yr$^{-1}$)   & 
10.9±0.3 yr ($\nu$=0.092±0.002  yr$^{-1}$)  \\ \hline
\end{tabular}
\end{table*}

\begin{figure*}
	\includegraphics[width=\textwidth]{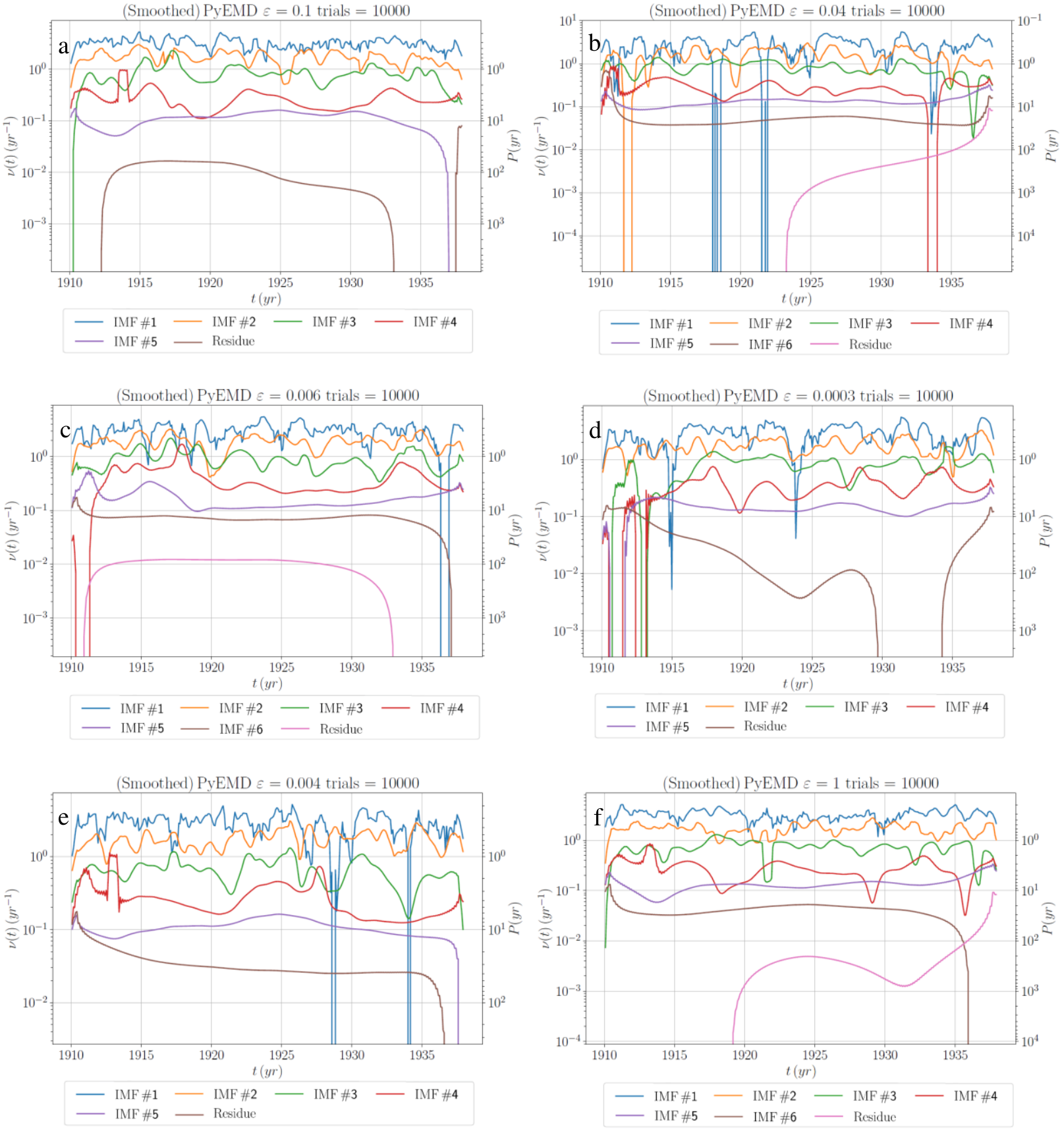}
    \caption{Instantaneous frequencies for all the $\widetilde{IMF}s$ and the residue for all the ANSAI related to the monthly sum of occurrences (left panels) and areas (right panels) of sunspot and solar plages groups during 1910—1937. The 2-set of panels \textit{a-b}, and \textit{c-d} are related respectively to sunspot data from the EO and RGO-USAF/NOAA. Panels \textit{e-f} are related to solar plage data from EO.}
    \label{fig:Figure7}
\end{figure*}

We also applied the CEEMDAN algorithm to the 1910--1937 EO and RGO-USAF/NOAA solar data. The evolution over time of the different instantaneous frequencies obtained from all the $\widetilde{IMFs}$ and the final residue is shown in Figure~\ref{fig:Figure7}. In this case, we fixed at 10000 the number of trials for not observing any changes between two independent runs. Thus, we found between 5 and 6 components depending on the considered solar structure and property. Again, $\widetilde{IMF}~\#$1 and $\#$2 also present very irregular periods, and shorter than the periodic signal \textit{E}, $\widetilde{IMFs}~\#$3 and $\#$4 present irregular periods, $\widetilde{IMF}~\#$5 present the most regular period, and $\widetilde{IMF}~\#$6 presents a regular period on the order of 15-25 yr.

The resulting averaged periods for all the $\widetilde{IMF}~\#$5 associated to each solar structure and property, shown in Table \ref{Table12}, suggest that the $\sim7.5$ yr periodicity is not a harmonic, and also underwrite our previous statements related to the slight differences between both solar structures considered.

\begin{table*}
\centering
\caption{Length of the mean periods obtained from averaging the $\widetilde{IMF}~\#$5 instantaneous frequency values via CEEMDAN algorithm for all the ANSAI related to the monthly sum of occurrences and areas of all sunspot and solar plages groups series during 1910--1937.}
\label{Table12}
\begin{tabular}{lll}
\hline
\textbf{Feature (observatory)} & \textbf{Analysed years} & \textbf{Mean period}                \\ \hline
Sunspot areas (OE)             & {[}1918—1929.5{]}       & 7.19±0.04 yr   ($\nu$=0.139±0.001 yr$^{-1}$) \\
Sunspot areas (RGO)            & {[}1918—1930{]}         & 7.14±0.07 yr ($\nu$=0.140±0.001 yr$^{-1}$)   \\
Solar plage areas (OE)         & {[}1918—1930{]}         & 7.68±0.06 yr ($\nu$=0.130±0.001 yr$^{-1}$)   \\
Sunspot occurrences   (OE)     & {[}1918—1927{]}         & 7.35±0.08 yr ($\nu$=0.136±0.002 yr$^{-1}$)   \\
Sunspot occurrences   (RGO)    & {[}1917.5—1936.5{]}     & 7.5±0.1 yr ($\nu$=0.124±0.002 yr$^{-1}$)     \\
Solar plage   occurrences (OE) & {[}1918—1931{]}         & 7.8±0.1 yr ($\nu$=0.128±0.002 yr$^{-1}$)     \\ \hline
\end{tabular}
\end{table*}

\subsection{Wavelet analysis}
\label{sec:Wavelet analysis}

In the light of the previous results, we analysed the divided sunspot series belonging to the RGO-USAF/NOAA by using the wavelet power spectrum \citep{Farge1992} in order to confirm whether the $\widetilde{IMF}~\#$5 suffers a drift in its value over the time, or if the instability of the oscillating component is caused by the presence of additional periodic signals in a similar time scale and the CEEMDAN algorithm is not able to separate them (mode mixing).  

The wavelet power spectrum, which is defined as the square of the CWT amplitude, i.e., $|CWT_n(s)|^2$ \citep{Torrence1998}, is a potent tool for analysing localised variations of the power within a time series (\cite{Torrence1998}; see also \url{https://github.com/chris-torrence/wavelets} for the Python implementation used in this section). Hence, the resultant scaleogram provides information about the exact location of the detected signals simultaneously in both time (x axis) and frequency/period (y axis) domains \citep{Chowdhury2015, Ravindra2021}.

The scaleograms corresponding to the twelve above-mentioned intervals of time related to sunspots groups recorded by the RGO-USAF/NOAA during 1874--2016 were computed considering their respective $\omega_0$, listed in Table~\ref{Table7}, and are available online as three figures of eight panels each (Figures S-4, S-5, and S-6). Left and right panels correspond to the scaleograms generated from the ANSAI related to the monthly sum of sunspots groups occurrences and areas, respectively. All the scaleograms suffer from edge-effect artefacts at both ends of each time series, which is shown in each scaleogram by dark hatched areas called cone of influence (COI). These edge effects, caused by having a finite-length time series, lead to a power reduction within the COI \citep{Chowdhury2016}. Thus, information in the scaleograms within their COI should be treated as suspect and is not highly significant. On the other hand, confidence levels were computed by following the procedure indicated in \citet{Auchere2016}, which is analogous to the above-mentioned case of TAWS (see Table~ \ref{methods}), but now using other $r$ and $s$ parameters, also empirically calculated in \citet{Auchere2016}. 

The obtained results confirm the presence of five different periodic signals in both occurrences and areas N-S asymmetry data. However, as shown in Section \ref{sec:Power spectrum analysis}, some of them present temporal discontinuities, and also show a drift in the period with time. Moreover, sometimes periodicities are not distinguishable since they overlap among them. In addition, the COI prevents a complete analysis of those signals with a period of 8 yr or greater. Notwithstanding the foregoing, we could extract some extra information of the periodicities as regards their temporal location. 

It can be seen that the signal with an associated period of  $\sim$1.5 yr is intermittent in intervals of 5-7 years approximately. This periodicity is statistically significant in areas especially during 1881--1883, 1945--1960 and 1978--1984, and in occurrences during 1916--1940.

With reference to the $\sim$4.0 yr periodic signal, it is is statistically significant in areas during 1892--1895, 1904--1909, 1935--1943, and 1958--1963. Regarding occurrences, the periodic signal is significant during 1892--1895, 1920--1937, 1958--1962, and 1983--1995.

Regarding the $\sim$5.5 yr periodicity, it is statistically significant in areas during 1882--1890, 1953--1958, and 1996--2000. In the case of occurrences, the periodic signal is significant during 1887--1890, 1897--1906, 1953--1958 and 1995--2007.

Finally, several contours centered at $\sim$9 yr and $\sim$13 yr can also be observed over the time, but they cannot be considered as statistically significant because they are located inside the COI.

\begin{figure*}
	\includegraphics[width=\textwidth]{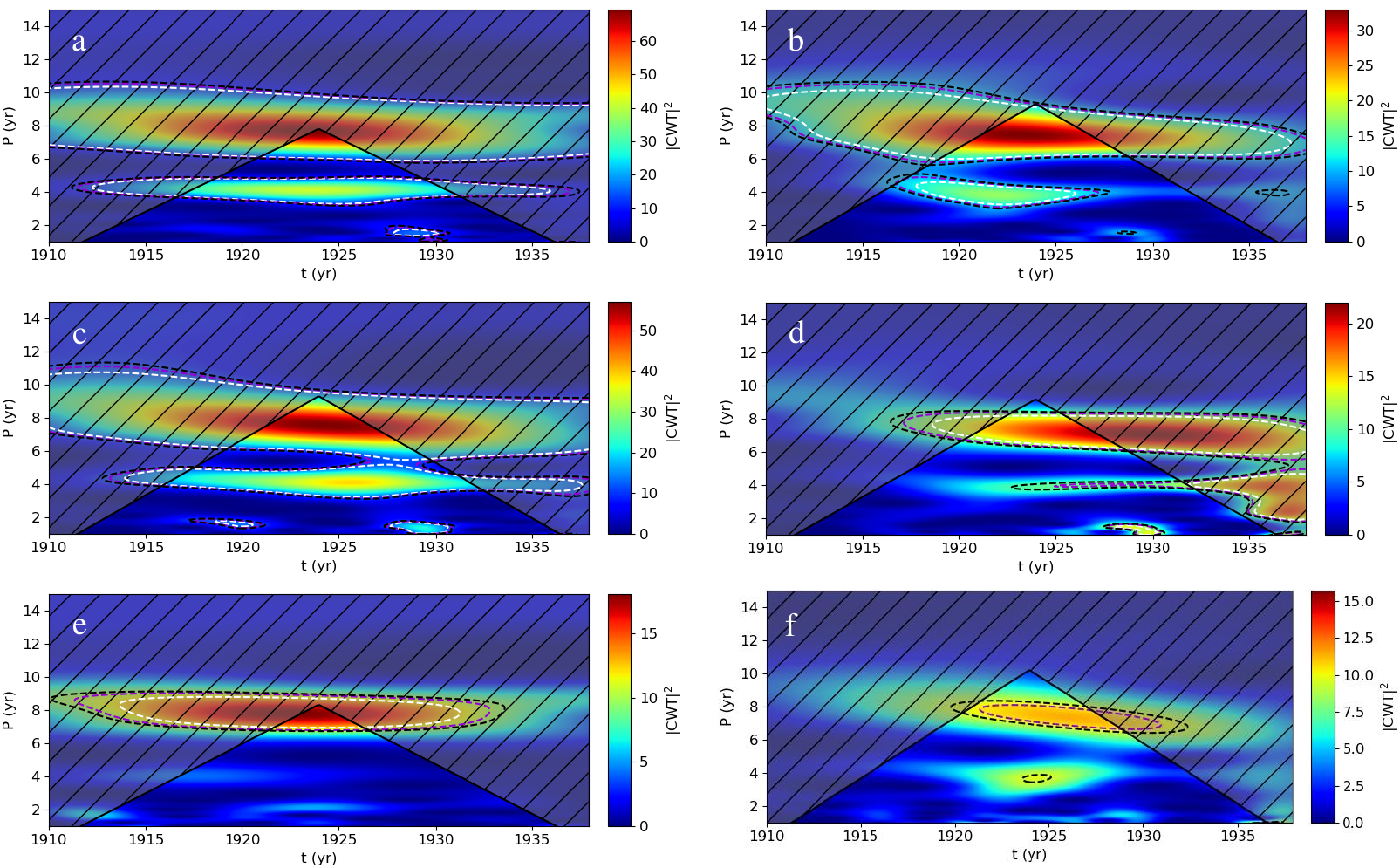}
    \caption{Scaleograms for all the ANSAI related to the monthly sum of occurrences (left panels) and areas (right panels) of sunspot and solar plages groups during 1910—1937. The 2-set of panels \textit{a-b}, and \textit{c-d} are related respectively to sunspot data from the EO and RGO-USAF/NOAA. Panels \textit{e-f} are related to solar plage data from EO. Dark hatched areas represent the COI. Dotted lines represent the confidence levels: $99\%$ white; $95\%$ violet; $90\%$ black.}
    \label{fig:Figure8}
\end{figure*}

We also computed the scaleograms generated from the ANSAI related to the monthly sum of sunspots and solar plage groups occurrences and areas recorded by the EO and RGO-USAF/NOAA during 1910--1937 (see Figure~\ref{fig:Figure8}). We considered the $\omega_0$ values which are listed in Table~\ref{Table2}. Two different periodicities detected around 4.0-4.2 yr and 7.0-7.9 yr in the spectral analysis can also be observed practically during the entirely time lapse, with the exception of solar plage occurrences. Moreover, the intermittent $\sim1.5~yr$ periodic signal is also present during 1917--1931, but only is statistically significant in sunspots. Finally, the slight differences observed in the power spectrum and CEEMDAN analyses between the different periods detected in both sunspot and solar plage data sets are also present in this technique.

\section{Discussion}

The Ebro Observatory solar series may contribute to broaden the knowledge about the periodic behaviour of the N-S asymmetry in solar activity in terms of the monthly sum of areas and occurrences of sunspots and solar plages groups, two structures located at the solar Photosphere and Chromosphere, respectively. For this purpose, we computed each N-S asymmetry indices time series related to these properties by taking the daily measurements associated to both solar structures, which can be found in our annual heliophysics bulletins during 1910--1937, both inclusive. 

After detecting a cyclic behaviour with an averaged period of $7.9 \pm 0.2~yr$ in the N-S asymmetry indices of the above-mentioned properties in both sunspots and solar plage groups during 1910--1937, we deeply analysed our time series and also examined the RGO-USAF/NOAA sunspot data series from 1874 to 2016 in order to verify our results and extend in time our study. Thus, we applied several techniques and mathematical tools to all time series, such as the power spectrum analysis, the CEEMDAN method or the wavelet analysis, and we found evidence of additional periodicities in different time scales.

The power spectrum, and in particular the Lomb-Scargle periodogram of each entire data set allowed us to quantify the period of each signal found. Thus, during 1910--1937, we obtained two statistically significant periodicities with means of 4.10±0.04 yr and 7.57±0.03 yr in both solar structures and properties analysed. Furthermore, during 1874--2016, we found three significant periodicities with mean values of 1.45 yr, 4.23±0.01 yr, and 8.95±0.03 yr in sunspots and also both properties considered. These three last periodicities were extended to five by dividing the full series into time intervals of three solar cycles each, obtaining the following averages: 1.47±0.02 yr \textit{(E)}, 3.83±0.06 yr \textit{(D)}, 5.4±0.2 yr \textit{(C)}, 9.0±0.2 yr \textit{(B)}, and 12.7±0.3 yr \textit{(A)}. Nevertheless, they seem to be weak since in general they are not statistically significant, and intermittent, since they are not present in all time intervals.

The CEEMDAN method results are quite consistent with those obtained by the power spectrum analysis for the time span of 1910—1937, and also for all datasets of three solar cycles within 1874—2016. Thus, the obtained CEEMDAN $\widetilde{IMF}~\#$1 and $\#$2 could be noise or true physical signals with periods smaller than those studied in this paper; $\widetilde{IMF}~\#$3 and $\#$4 could be associated respectively to periodicities \textit{E} ($\sim$1.5 yr) and \textit{D} ($\sim$4.0 yr), despite the fact that they present a very irregular period, which is probably due to mode mixing between these modes; $\widetilde{IMF}~\#$5 could present mode mixing of periodicities \textit{B} (9.0 yr) and \textit{C} ($\sim$5.5 yr); $\widetilde{IMF}~\#$6 could be associated to periodicity \textit{A} ($\sim$13.0 yr), but is strongly affected by the mode mixing phenomenon with other periodic signals, and $\widetilde{IMF}~\#$7 could reveal the existence of periodicities in time scales larger than the 11-years solar cycle, as shown by \citet{Ballester2005}, who found a periodicity of 43.25 yr in the ANSAI related to the monthly sunspot areas recorded by the RGO-USAF/NOAA during 1874--2004. Notwithstanding the above, it can be seen that the CEEMDAN method only provides conclusive evidence of the presence of the 7.0-7.9 yr periodicity, while the other periodic variations detected with the other methods are not clearly revealed by the CEEMDAN. We suggest that mode mixing between different periodic components is the cause for this CEEMDAN failure and that the 7.0-7.9 yr periodicity has a large enough amplitude that prevents mode mixing from distributing this signal among different $\widetilde{IMF}$s. Hence, a single $\widetilde{IMF}$ can catch its variability, although we have seen that this $\widetilde{IMF}$ also suffers from mode mixing. Furthermore, thanks to the CEEMDAN method properties we confirmed that all periodic signals detected were not harmonics among them. 

Finally, the temporal spans of all periodicities within the 11th-24th solar cycles, and also during 1910--1937 are determined from the wavelet analysis. Thus, their intermittence and strength over the time can be studied. Again, all results are consistent with the preceding ones, and in agreement with \citet{Bazilevskaya2014} statements. They concluded in their investigation on QBOs that despite being highly irregular and resembling a set of intermittent pulses/waves with signatures of stochasticity, QBOs are more prominent during periods of high solar activity. The most evident example of the preceding observations is presented by the periodicity \textit{E} ($\sim$1.5 yr), since it is especially strong during the ascending phase and maximum of the vast majority of solar cycles. Nevertheless, it seems to be absent during the 15th and 23rd solar cycles. The clear intermittence and also the strength which presents the $\sim$1.5 yr periodicity during the active phase of the Sun has been reported in different solar indices \citep{Krivova2002,Cadavid2005,Vecchio2009,Simoniello2013,Chowdhury2016,Chowdhury2019}. 

We compared the five periodicities found in our study during 1874--2016 with others which have been observed in previous investigations of different solar indices by using several methods of analysis. We summarised all the comparable results in Table~\ref{Table13}. It is important to remark that all the periodicities in Table~\ref{Table13} are related to features or structures which take place within the different layers of the solar atmosphere. However, these periodicities can also be transferred to the heliosphere through the open magnetic flux \citep{Lockwood2001,Bazilevskaya2014}, and reflected in several phenomena as the velocity of solar wind \citep{Richardson1994, Li2017}. Hence, the study and characterisation of these periodicities may be useful in the space weather field \citep{Bazilevskaya2014}.

\begin{table*}
\centering
\caption{Selected papers related to periodicities in different solar indices, which could be associated to periodicities \textit{A}, \textit{B}, \textit{C}, \textit{D}, and E found in our study. }
\label{Table13}
\begin{tabularx}{\textwidth}{XXXXXX}
\hline
\textbf{Author(s)} &
  \textbf{Solar index} &
  \textbf{Data source} &
  \textbf{Analysed years} &
  \textbf{Method of analysis} &
  \textbf{Periodicity (location)} \\ \hline
\multicolumn{6}{c}{\textit{Periodicity A (12.7±0.3 yr)}} \\
\citet{Knaack2004} &
  NNSAI (monthly sunspot areas) &
  RGO-USAF/NOAA &
  1976--2004 &
  Wavelet &
  13 yr (1960—1985) \\
\citet{Javaraiah2020} &
  NNSAI (monthly sunspot areas) &
  RGO and Debrecen &
  1874--2017 &
  Wavelet (Morlet) &
  12.8 yr (1880--1920, 1960--1990) \\ \hline
\multicolumn{6}{c}{\textit{Periodicity B (9.0±0.2 yr)}} \\
\citet{Duchlev1996} &
  NNSAI (yearly long-lived filaments) &
  Meudon &
  1874--2003 &
  Power spectrum (DFT) &
  8.75 yr \\
\citet{Krivova2002} &
  Monthly averaged sunspot areas &
  RGO &
  1874—1976 &
  Power spectrum (GWS) + Wavelet (Morlet) &
  8.99 yr \\
\citet{Ballester2005} &
  ANSAI (monthly sunspot areas) &
  RGO-USAF/NOAA &
  1874--2004 &
  Power spectrum (LSP) &
  8.65 yr \\
\citet{Zolotova2007} &
  NNSAI (monthly sunspot areas) &
  RGO-USAF/NOAA &
  1874--2003 &
  Power spectrum (DFT) &
  8.64-9.26 yr \\
\citet{Javaraiah2020} &
  NNSAI (monthly sunspot areas) &
  RGO and Debrecen &
  1874--2017 &
  Wavelet (Morlet) &
  9 yr (1920--1950; 1990-2017) \\ \hline
\multicolumn{6}{c}{\textit{Periodicity C (5.4±0.2 yr)}} \\
\citet{Krivova2002} &
  Monthly averaged sunspot areas &
  RGO &
  1874—1976 &
  Power spectrum (GWS) + Wavelet (Morlet) &
  5.23 yr \\
\citet{Chowdhury2016} &
  Plage index (yearly average) &
  Kodaikanal &
  1907--1998 &
  Power spectrum (DFT and Maximum Entropy) and Wavelet (Morlet) &
  5-5.5 yr (1937--1991) \\
\citet{Zhu2018} &
  Daily New International Sunspot Numbers &
  WDC-SILSO &
  1818--2014 &
  Power spectrum (LSP) &
  5.445 yr \\
\citet{Ravindra2021}&
  ANSAI (monthly averaged sunspot areas) &
  Kodaikanal &
  1921--2011 &
  Wavelet (Morlet) &
  5 yr (Solar Cycles 18-20, 22-23) \\ \hline
\multicolumn{6}{c}{\textit{Periodicity D (3.83±0.06 yr)}} \\
\citet{Knaack2004} &
  NNSAI (solar magnetic field) &
  Kitt Peak &
  1975--2003 &
  Power spectrum (DFT) + Wavelet &
  3.6 ± 0.3 yr (1978—1995) \\
\citet{Knaack2004}  &
  NNSAI (monthly sunspot areas) &
  RGO-USAF/NOAA &
  1976--2003 &
  Power spectrum (DFT) + Wavelet &
  3.9 yr (1982—1992) \\
\citet{Joshi2004} &
  NNSAI (daily soft X-ray flare index) &
  GOES data &
  1975--2003 &
  Power spectrum (LSP) &
  3.72 yr \\
\citet{Kolotkov2015} &
  Helioseismic frequency shift &
  BiSON &
  1985--2014 &
  EEMD &
  3.42 yr \\
\citet{Roy2020} &
  ANSAI (daily Solar-Flare Index) &
  Kandilli &
  1976--2018 &
  Power spectrum (Date-compensated DFT) &
  3.78 yr \\ \hline
\multicolumn{6}{c}{\textit{Periodicity E (1.47±0.02 yr)}} \\
\citet{Ichimoto1985} &
  Daily number of H$\alpha$ flares &
  Tokyo Observatory &
  1965--1984 &
  Power spectrum (Maximum Entropy) &
  1.42 yr \\
\citet{Knaack2004}  &
  NNSAI (solar magnetic field) &
  Kitt Peak &
  1975--2003 &
  Power spectrum (DFT) + Wavelet &
  1.50 ± 0.04 yr (1978—1984) \\
\citet{Joshi2004} &
  NNSAI (daily soft X-ray flare index) &
  GOES data &
  1975--2003 &
  Power spectrum (LSP) &
  1.51 yr \\
\citet{Ballester2005} &
  ANSAI (monthly sunspot areas) &
  RGO-USAF/NOAA &
  1874--2004 &
  Power spectrum (LSP) &
  1.44 yr \\
\citet{Qiang2007} &
  Monthly averaged sunspot numbers &
  NOAA &
  1894--2003 &
  EMD &
  1.3-1.4 yr \\
\citet{Chowdhury2016} &
  Plage index (yearly average) &
  Kodaikanal &
  1907--1998 &
  Power spectrum (DFT and Maximum Entropy) and Wavelet (Morlet) &
  1.5-1.7 yr (1917--1920); 1.1-1.4 yr (1922--1933); 1.3-1.4 yr   (1934--1938); 1.3-2.4 yr (1947--1963); 1.25-1.5 yr (1967--1973); 1.2-2.4 yr   (1977--1993) \\ \hline
\end{tabularx}
\end{table*}

Another interesting point is the nature of the 7.0-7.9 yr periodic signal, found in all sunspots and solar plages asymmetry datasets during 1910--1937. This periodicity decreases its strength in the power spectrum to the extent that goes unnoticed when we consider the full series of RGO-USAF/NOAA (1874--2016), as shown in both panels in Figure~\ref{fig:Figure5}. This fact together with wavelet results suggest that it could be an independent signal which only exists approximately during 1917--1947, as can be seen in Figures S-4, S-5, and S-6, and in Figure~\ref{fig:Figure8}. Even taking into account the possibility that the 7.0-7.9 yr periodicity is actually a harmonic, it only could be as a result of a linear combination of the periodic signals \textit{A}, \textit{B}, \textit{C}, \textit{D} and \textit{E}. But during the 15th, 16th and 17th solar cycles, the 7.0-7.9 yr periodic signal is present and however, signals \textit{A}, \textit{B} and \textit{C} are missing. Thus, the only possible way to obtain this periodicity is from signals \textit{D} ($\sim$4 yr) and \textit{E} ($\sim$1.5 yr). But as it can be seen in all panels of Figure \ref{fig:Figure4}, the 7.0-7.9 yr periodicity is slightly stronger than signal \textit{D} and far stronger than signal \textit{E}, which contradicts the initial assumption. Moreover, the hypothesis that this periodicity exists as a drift of the periodic signal \textit{B} ($\sim$9.0 yr) could be contradicted since during the 16th, 17th and 18th solar cycles, the two periodicities exist simultaneously. Finally, the 7.0-7.9 yr periodic signal could not be associated to the periodicity \textit{C} ($\sim$5.5 yr) drift since during the 16th, 17th and 18th solar cycles both periodicities also exist simultaneously. In addition to our finding, other authors also detected similar periodicities in other indices of the solar activity, although none of them is related to the N-S asymmetry. \citet{Krivova2002} found a 7.89 yr periodicity by analysing the GWS originated from the RGO monthly sunspot areas data during 1874--1976, and also a 7.97 yr period by analysing the monthly values of the Zürich relative sunspot numbers during 1749--2001. In addition, \citet{Zhu2018} found a peak of 7.990 yr in the LSP associated to the daily New International Sunspot Numbers during 1818--2014. Moreover, \citet{Chowdhury2019} examined the monthly sunspot number taken at Kanzelhöhe Observatory during 1944--2017. They found a significant period of approximately 7 years in both solar hemispheres data by employing the multitaper period analysis and the Morlet wavelet transform analysis. 

Moreover, the 4.0-4.2 yr periodicity found in all sunspots and solar plages asymmetry datasets during 1910--1937, could be associated to the periodic signal \textit{D}.

The spectral analysis of the periodicities found during 1910--1937 combined with the results obtained by using the CEEMDAN and wavelet techniques reveal that solar plage groups present slightly larger periods in both areas and occurrences in comparison with sunspots. In spite of the fact that sunspots and solar plages are two solar structures magnetically coupled, this result manifests that both Photosphere and Chromosphere present small differences in terms of the observed solar activity. In addition, the Pearson’s correlation coefficients calculated between the different asymmetry indices of sunspot and solar plage groups reflect these differences. Similar conclusions with regard to solar rotation and to the N-S asymmetry related to the occurrences and areas of sunspots and solar plages were extracted in \citet{dePaula2016} and in \citet{dePaula2020}, respectively. Hence, we strongly think that this issue should be considered in further studies. 

Even though there is no certain explanation to the origin of all these periodicities yet, the vast majority of authors suggest that all detected Rieger periodicities and QBOs have their origin in the solar interior. Nevertheless, two main hypotheses have been formulated. The first one suggests the existence of two solar dynamos operating at different depths, one at the base of the convection zone and another situated near the solar surface \citep{Benevolenskaya1998,Bazilevskaya2014,Obridko2014,Beaudoin2016,Zharkova2017,Chowdhury2019,Ravindra2021}. The second hypothesis is in terms of magnetic Rossby waves in the solar tachocline \citep{Lou2003,Knaack2005,Chowdhury2009,Zaqarashvili2010,Zaqarashvili2011,Bazilevskaya2014,Gurgenashvili2016,Gurgenashvili2017}. These Rossby waves belong to a subset of global tidal waves (which are described by Laplace’s tidal equation) that can exist in a fluid on the surface of a rotating sphere or on a thin layer, such as the solar tachocline \citep{Lou2000,Chowdhury2019}. As argued by \citet{Gurgenashvili2016} and \citet{Ravindra2021}, the periodicities might correspond to different harmonics of magnetic Rossby waves.

A third explanation which is steadily gaining recognition is the following. Several authors as \citet{Grandpierre1996, Abreu2012, Okhlopkov2016,Scafetta_2016}, or \citet{Javaraiah2020} suggested that some specific configurations of the giant planets (Jupiter, Saturn, Uranus and Neptune) could influence the solar differential rotation rate and hence, the strength of the solar dynamo. In particular, \citet{Javaraiah2020} showed that 12-13-yr periodicity in the NNSAI related to the monthly sunspot areas obtained from Greenwich and Debrecen Observatories during 1874--2017 occurred during approximately the same times as the corresponding periodicity in the mean absolute difference of the orbital positions $(\psi_D=\psi_i-\psi_j)$, i.e., ecliptic longitudes $\psi_i$, of the giant planets. Hence, new solar dynamo models which includes the influence of the planetary tidal forces on the solar activity has been proposed \citep{Stefani_2016,Stefani_2019}.

Thus, to better understand the physical processes behind all these periodicities, more investigation on the Sun’s internal dynamics, and on the gravitational influence of the solar giant planets is required. In any event, the research on the different solar indices, and especially on the distribution of photospheric sunspots and chromospheric solar plages may be useful to analyse the distribution of magnetic fields in the solar interior and it also provides strong observational constraints on the solar dynamo theory \citep{Berdyugina2003}.

\section{Conclusions}

In this work, we examined the historical heliophysics catalogues from the Ebro Observatory in the period 1910--1937, which contain daily records of sunspot and solar plage groups, two structures located at different layers of the solar atmosphere. Our study was focused on the periodic behaviour of the N-S asymmetry in the monthly sum of areas and occurrences of these structures observed during 1910--1937, which initially appeared to have a period of $7.9 \pm 0.2~yr$, depending on the solar structure and property considered. In order to obtain more information about this phenomenon, which has not been reported in the bibliography yet, and study its prevalence in time, we expanded the initial temporal window of our study to 1874--2016 by using sunspot data recorded by the RGO-USAF/NOAA. Thus, an extensive analysis of the data was done, by using several methods and mathematical tools, such as the power spectrum analysis (DFT, TAWS and LSP), the CEEMDAN algorithm, and the wavelet analysis. From our findings, we highlight the following results:

\begin{enumerate}
\item The ANSAI in the monthly sum of areas and occurrences of sunspot and solar plage groups during 1910--1937 presents two statistically significant periodic signals with an associated period of 4.10±0.04 yr and 7.57±0.03 yr, respectively. 

\item The analysis of the ANSAI data belonging to the monthly RGO-USAF/NOAA sunspot groups records during 1874--2016 allowed us to find out the periodic behaviour in five different time scales. In both areas and occurrences, we detected two practically stable periodicities of 1.47±0.02 yr and 3.83±0.06 yr, respectively. It is also important to remark the intermittent behaviour of the 1.47±0.02 yr periodicity every 5-7 years, present during the ascending phase and maximum of all solar cycles, except the 15th and 23rd. All in all, these two periodic signals coexist with other three additional, but much more discontinuous components with periods of 5.4±0.2 yr, 9.0±0.2 yr, and 12.7±0.3 yr, respectively. The absence of solar plage data in our data base outside the time span of 1910--1937 did not allow us to determine if these solar structures also present those periodic signals in the considered properties.

\item Despite most of the periodicities during 1874--2016 are found below the 90$\%$ confidence level, their persistence in time and their simultaneous presence in both areas and occurrences suggests they are not fruit of randomness. Moreover, at least during the 15th, 16th and 17th solar cycles, all the detected periodicities are statistically significant, and their periods are consistent with those found by analysing the sunspot and solar plage EO data during 1910--1937, and also with those found by other authors in several indices of the solar activity. 

\item The 7.0-7.9 yr periodicity could not be an artefact but a real signal in the solar N-S asymmetry and only its fluctuating apparition along the time can explain why it evaded its manifestation in the analysis of other preceding works. Future solar models should integrate this period.

\item At least during 1910--1937, solar plage groups present slightly larger values of the period in both periodic signals with respect to sunspot groups. Despite sunspots and solar plages are magnetically coupled, the previous statement together with a fairly low correlation between the asymmetry indices related to these two structures could be an evidence of small differences in the solar activity within the different layers of the solar atmosphere, and must be considered in further studies. 

\end{enumerate}

\section*{Acknowledgements}

This work would not have been possible without the efforts of the Ebro Observatory observers and the people who digitised our heliophysics bulletins at the beginning of the 21st century. We also would like to express our gratitude to the NASA’s Marshall Space Flight Center Solar Physics Group, who provide good quality open access data in their portal, as well as all the RGO, USAF and NOAA staff, who made possible their sunspot series. We also acknowledge enlightening discussions with F. Auchère about the calculation of the mean power spectrum of the DFT and thank M. Kriginsky for bringing to our attention the PyEMD library, used to perform the CEEMDAN analysis. The TAWS, the CWT and their significance have been computed with the Python wavelet software provided by E. Predybaylo based on Torrence and Compo (1998) and is available at URL: \url{https://github.com/chris-torrence/wavelets}. This publication is part of the R+D+i project PID2020-112791GB-I00, financed by MCIN/AEI/10.13039/501100011033. Finally, we thank the anonymous reviewer for very useful comments that helped improve this manuscript.

\section*{Data Availability}

The data underlying this article are available in the Ebro Observatory website: \url{http://www.obsebre.es/en/observatori-publications}, and in the Solar Physics Group at NASA’s Marshall Space Flight Center website: \url{https://solarscience.msfc.nasa.gov/greenwch.shtml}.



\bibliographystyle{mnras}
\bibliography{example} 








\bsp	
\label{lastpage}
\end{document}